\documentclass[apj]{emulateapj}

\usepackage{natbib}
\usepackage{color}

\usepackage{bm}
\usepackage{lscape}

\newcommand{\eg}{e.g., }
\newcommand{\ie}{i.e., }
\newcommand{\Msun}{M_{\odot}}
\newcommand{\kms}{km~s$^{-1}$}

\newcommand{\Cofs}{$^{56}$Co}
\newcommand{\Nifs}{$^{56}$Ni}

\newcommand{\Mej}{M_{\rm ej}}
\newcommand{\KE}{E_{\rm K}}
\def\gsim{\mathrel{\rlap{\lower 4pt \hbox{\hskip 1pt $\sim$}}\raise 1pt
\hbox {$>$}}}
\def\lsim{\mathrel{\rlap{\lower 4pt \hbox{\hskip 1pt $\sim$}}\raise 1pt
\hbox {$<$}}}

\def\ion#1#2{{\rm #1}~{\sc #2}}

\shorttitle{Radiative Transfer Simulations for NS Merger Ejecta}
\shortauthors{Tanaka \& Hotokezaka}

\begin{document}

\title{
Radiative Transfer Simulations for Neutron Star Merger Ejecta}
\author{
Masaomi Tanaka\altaffilmark{1} and
Kenta Hotokezaka\altaffilmark{2}
}

\altaffiltext{1}{National Astronomical Observatory of Japan, Mitaka, Tokyo, Japan; masaomi.tanaka@nao.ac.jp}
\altaffiltext{2}{Department of Physics, Kyoto University, Kyoto, Japan; hotoke@tap.scphys.kyoto-u.ac.jp}

\begin{abstract}
The merger of binary neutron stars (NSs) is among 
the most promising gravitational wave (GW) sources. 
Next-generation GW detectors are expected to detect 
signals from the NS merger within 200 Mpc. 
Detection of electromagnetic wave (EM) counterpart 
is crucial to understand the nature of GW sources. 
Among possible EM emission from the NS merger, 
emission powered by radioactive r-process nuclei 
is one of the best targets for follow-up observations.
However, prediction so far does not take into account 
detailed r-process element abundances in the ejecta.
We perform radiative transfer simulations for 
the NS merger ejecta including all the r-process elements 
from Ga to U for the first time.
We show that the opacity in the NS merger ejecta is 
about $\kappa = 10\ {\rm cm^2\ g^{-1}}$, which is higher than 
that of Fe-rich Type Ia supernova ejecta by a factor of $\sim$ 100.
As a result, the emission is fainter and longer than previously expected.
The spectra are almost featureless due to the high expansion velocity
and bound-bound transitions of many different r-process elements.
We demonstrate that the emission is brighter for 
a higher mass ratio of two NSs 
and a softer equation of states adopted in the merger simulations.
Because of the red color of the emission, 
follow-up observations in red optical and near-infrared (NIR) wavelengths 
will be the most efficient.
At 200 Mpc, expected brightness of the emission is 
$i=$22-25 AB mag, $z=$21-23 AB mag, and 21-24 AB mag in NIR $JHK$ bands.
Thus, observations with wide-field 4m- and 8m-class optical telescopes 
and wide-field NIR space telescopes are necessary.
We also argue that the emission powered by radioactive energy
can be detected in the afterglow of nearby short gamma-ray bursts.
\end{abstract}


\section{Introduction}
\label{sec:intro}

The merger of binary neutron stars (NSs) 
is among the most promising candidates
for the direct detection of gravitational waves (GWs).
Next-generation GW detectors, such as advanced 
LIGO, advanced VIRGO, and KAGRA
\citep{abadie10,kuroda10,accadia11,ligo13},
are expected to detect GWs from the NS merger at 
a distance within $200$ Mpc.
Statistical studies have shown that 
the number of GW detection will be 
in a range of 0.4-400 per year \citep{abadie10rate,coward12}.

Follow-up observations of electromagnetic wave (EM) counterparts
are essentially important.
Only with the GW detection, 
the position of the sources can only be moderately determined
with a localization of about 10-100 deg$^2$ \citep[\eg][]{ligo12,ligo13,nissanke13}.
Therefore, to fully understand the nature of the GW sources,
EM observations should pin down the position of the sources
and identify the host galaxy and environment.
%and accurately determine the distance to the sources.

Possible EM emissions from the NS merger 
\citep[\eg][]{kochanek93,metzger12,rosswog13,piran13} are
(1) short gamma-ray bursts (GRBs),
(2) radio/optical afterglow, 
and (3) emission powered by the radioactive decay energy. 
Among them, the last one is of great interest
because of the isotropic nature of the emission and a relatively 
short time delay after the merger (\ie the detection of GWs).

By the merger of binary NSs, 
a small fraction of matter is expected to be ejected 
\citep[\eg][]{rosswog99,rosswog05,lee07,duez10,goriely11,rosswog13a,hotokezaka13,bauswein13}.
Hereafter we call this ejected material ``NS merger ejecta''.
NS merger ejecta is one of the promising sites
for r-process nucleosynthesis 
\citep[\eg][]{lattimer74,lattimer76,eichler89,freiburghaus99,roberts11,goriely11,korobkin12,bauswein13}.
Some of the synthesized r-process nuclei
can provide radioactive decay energy with a timescale of 1-10 days or so.
\citet{li98} first proposed that 
this radioactive decay energy gives rise to 
the emission in the UV-optical-IR (UVOIR) wavelength range.
This emission, which is is similar to 
the emission of supernovae (SNe) powered by \Nifs, 
has been called 
``macronova'' \citep{kulkarni05}, 
``kilonova''\citep{metzger10,metzger12}, or ``mini-SN''.

The brightness and the duration of the emission are 
mainly determined 
by (1) the mass and (2) the velocity of the ejecta,  
and also (3) opacity in the ejecta.
\citet{metzger10} first presented detailed study of 
this emission by taking into account the radioactive energy 
based on the nucleosynthesis calculations.
\citet{roberts11} and \citet{goriely11} also 
showed expected emission powered by radioactivity 
using their hydrodynamic and nucleosynthesis calculations.
However, since the opacity in the r-process element-rich ejecta
was poorly known, 
they have assumed that the opacity is similar to 
that of Fe-rich Type Ia SNe, \ie
$\kappa \sim 0.1\ {\rm cm^2\ g^{-1}}$ \citep{pinto00}.

Recently, \citet{kasen13} and \citet{barnes13}
evaluated the opacity of a few representative lanthanoid elements.
They showed that the opacity of these elements are 
higher than that of Fe by a factor of about 100.
They showed that the expected emission becomes 
fainter and longer than previously expected, 
as a result of the high opacity.

In this paper, we perform radiative transfer simulations
for NS merger ejecta including {\it all} the r-process elements 
for the first time.
For this purpose, we build a new line list for r-process elements
from VALD database \citep{piskunov95,ryabchikova97,kupka99,kupka00}, 
including about 100,000 bound-bound transitions. 
This strategy is complementary to that taken by \citet{kasen13} and 
\citet{barnes13}, 
who constructed detailed models of a few lanthanoid elements.

We first describe the details of a newly-developed, 
three-dimensional (3D), time-dependent, and multi-frequency 
Monte Carlo (MC) radiative trafer code in Section \ref{sec:code}.
Models for the NS merger ejecta are presented in Section \ref{sec:models}.
We present results of radiative transfer simulations 
in Sections \ref{sec:simple}.
We show that the opacity for the mixture of r-process elements
is as high as $\kappa = 10\ {\rm cm^2\ g^{-1}}$,
which is consistent with the results by \citet{kasen13,barnes13}.
We also show that the spectral features are smeared out 
by bound-bound transitions of many different r-process elements.
In Section \ref{sec:realistic},  
we demostrate that the EM emission depends on 
the mass ratio of the binary NSs, and also on 
the equation of states (EOSs) adopted in the merger simulations.
Based on these results, strategy for 
EM follow-up observations are discussed in Section \ref{sec:implications}.
We also argue that the emission can be possibly detected in the 
afterglow of nearby short GRBs.
Finally, we give conclusions in Section \ref{sec:conclusions}.

%%%%%%%%%%%%%%%%%%%%%%%%%%%%%%%%%%%%%%%%%%%%%%%%%%%% 
% Figure  %%%%%%%%%%%%%%%%%%%%%%%%%%%%%%%%%%%%%%%%%% 
%%%%%%%%%%%%%%%%%%%%%%%%%%%%%%%%%%%%%%%%%%%%%%%%%%%% 
\begin{figure*}
\begin{center}
\includegraphics[scale=1.2]{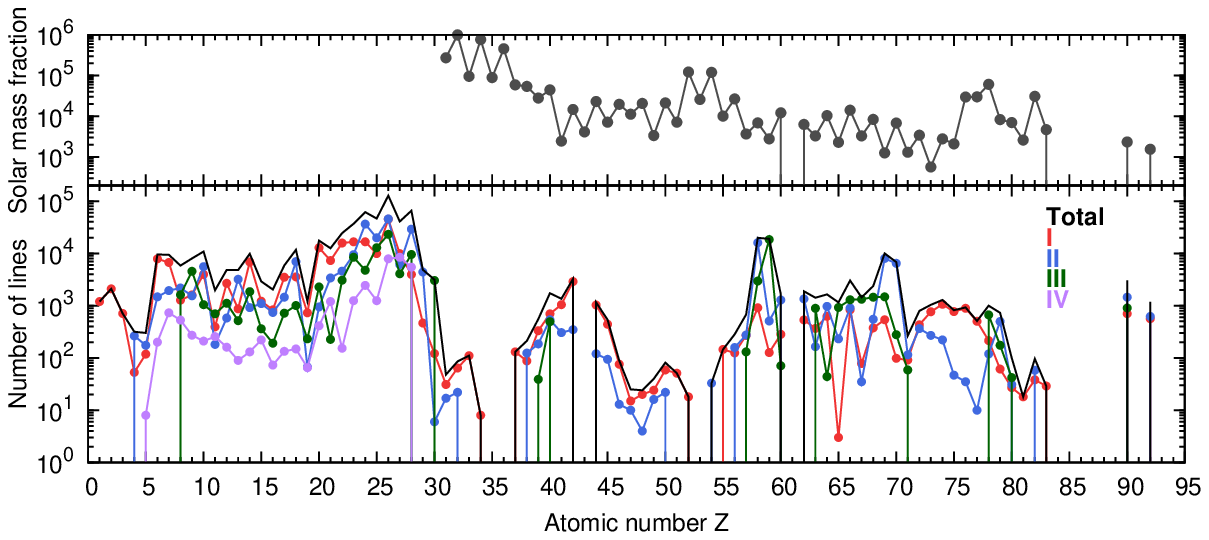}
\caption{
({\it Upper}) Solar abundance r-process abundance ratio in mass fraction
\citep{simmerer04}. 
The abundance is normalized with $X$(Ge) = $10^6$.
({\it Lower}) The number of bound-bound transition data for different elements.
Different colors show different ionization states, 
from neutral ({\sc I}) to triply ionized ({\sc IV}) ions.
The atomic data at $Z \le 30$ are taken from \citet{kurucz95} 
while the data at $Z \ge 31$ are compiled using the VALD database
\citep{piskunov95,ryabchikova97,kupka99,kupka00}.
It is shown that there is no data for triply ionized ions 
({\sc IV}, purple) at $Z \ge 31$.
}
\label{fig:nline}
\end{center}
\end{figure*}
%%%%%%%%%%%%%%%%%%%%%%%%%%%%%%%%%%%%%%%%%%%%%%%%%%%% 

\section{Radiative Transfer Code}
\label{sec:code}

\subsection{Overview}
\label{sec:overview}

We have developed a new 3D, time-dependent, 
multi-frequency radiative transfer code.
The code can be applied both for NS merger ejecta and SN ejecta.
For a given density structure and an abundance distribution,
the code computes the time series of spectra 
in the UVOIR wavelength range.
By integrating the spectra at a certain wavelength range 
with appropriate filter response curves, 
multi-color light curves are also computed.
To solve radiative transfer, 
we adopt MC technique following \citet{lucy05}.
The details of the code are presented in the subsequent sections.
Here we briefly summarize the procedures of simulations. 

After setting up the computational domain (Section \ref{sec:setup}),
photon packets are created 
by taking into account the radioactive 
decay (\Nifs\ for SNe and many r-process nuclei for NS mergers,
Section \ref{sec:packets}).
For SNe, $\gamma$-ray transfer is solved under the gray approximation
(Section \ref{sec:gamma}).
The absorbed $\gamma$-ray packets are converted into 
UVOIR packets (or so called $r$-packets by \citealt{lucy05}).
Transport of UVOIR packets is computed by taking into account
the electron scattering, and free-free, bound-free 
and bound-bound transitions (Section \ref{sec:UVOIR}).

By appropriately propagating the packets (Section \ref{sec:events}),
the temperature structure is determined
based on the photon flux (Section \ref{sec:temperature}).
By using the new temperature structure, 
ionization and excitation conditions are computed 
under the assumption of local thermodynamic equilibrium
(LTE, Section \ref{sec:ionization}).
Then, the UVOIR opacity is updated and MC transport is computed again.
Since the temperature is not known at first, 
these iterative calculations are performed at each time step.
Finally, escaping packets are counted,
which naturally give the time series of UVOIR spectra 
(Section \ref{sec:observations}).

The validity of our code is tested by comparing 
our results with those with other numerical codes.
First, a gray UVOIR transfer is tested 
for a simple model of Type Ia SN by \citet{lucy05}.
Then, the multi-frequency transfer is tested with 
the standard W7 model for Type Ia SN \citep{nomoto84}.
For both cases, we confirm a good agreement with the results with
3D MC gray transfer code by \citet{lucy05},
3D MC multi-frequency codes by \citet{kasen06} and by \citet{kromer09},
and one-dimensional (1D) multi-frequency code 
by \citet{blinnikov98,blinnikov00}.
Results of the test calculations are presented in Appendix A.

\subsection{Setup of computational domain}
\label{sec:setup}

A density structure and an abundance distribution 
are mapped into 3D Cartesian grid.
We typically use $32^3$ cells.
Thanks to the nearly homologous expansion of SN ejecta and NS merger ejecta 
(each fluid element expands in the radial direction
as $r = vt$, where $t$ is the time after the explosion 
or after the merger), 
we use velocity as a spatial coordinate.
The use of velocity coordinate has a great advantage
since we treat a wide range of time,
spanning about 2 orders of magnitude.
For a typical NS merger model, a spatial resolution is
$\Delta v \sim $ 2000 \kms\
($\Delta v \sim $ 1000 \kms\ for a typical SN model).
We do not consider any back reaction from radiation to hydrodynamics.
Thus, the density in each cell is simply updated 
as $\rho \propto t^{-3}$.

For the time grid, we use a logarithmically-spaced time step.
Simulations are performed typically from 0.1 days to 30 days 
with a time step of $\Delta \log (t/{\rm day}) = 0.02$ for NS merger models
(from 2 days to 50 days for SN models).
For the frequency grid, we use a linearly-spaced 
grid in the wavelength,
typically $\lambda =$ 100 - 25000 \AA\ with $\Delta \lambda = 10$ \AA.
At the center of the optical wavelength ($\sim 6000$ \AA), 
this wavelength resolution corresponds to the velocity resolution
of $\Delta v \simeq 500$ \kms.

\subsection{Creation of packets}
\label{sec:packets}

For MC radiative transfer, we use energy packets 
having equal energy \citep[\eg][]{lucy05}.
First, the total radioactive energy $E_{\rm rad}$ is 
divided into $N$ packets, so that
each packet has an equal comoving-frame energy 
$\epsilon_0 = E_{\rm rad}/N$.

For the case of SNe,
each packet is created as a $\gamma$-ray packet.
According to the total energy release by \Nifs\ decay ($E_{\rm Ni}$) and
\Cofs\ decay ($E_{\rm Co}$), each packet is designated as that from 
\Nifs\ or \Cofs; a fraction $E_{\rm Ni}/E_{\rm rad}$ is 
a \Nifs\ packet
while $E_{\rm Co}/E_{\rm rad}$ is a \Cofs\ packet.
If a packet is a \Nifs\ packet, the decay time is assigned by 
$t_{\rm active} = -t_{\rm Ni} \ln z$, where $t_{\rm Ni}$ 
is the lifetime of \Nifs.
Hereafter, $z$ is a random number from 0 to 1.
Similarly, if a packet is a \Cofs\ packet, 
$t_{\rm active} = -t_{\rm Ni} \ln z_1 - t_{\rm Co} \ln z_2$, 
where $t_{\rm Co}$ is the lifetime of \Cofs\
($z_1$, and $z_2$ are independent random numbers from 0 to 1).
These $\gamma$-ray packets are activated once the computation
reaches at $t > t_{\rm active}$.
An isotropic direction in comoving frame is also assigned
for each packet.

For the heating by many r-process radioactive nuclei in 
the NS merger ejecta, 
\citet{metzger10} showed the total radioactive power follows $t^{-1.2}$
(see also \citealt{korobkin12}).
Thus, a time of activation of each packet is assigned 
as $t_{\rm active} = t_{\rm 0, decay} z^{-5}$,
so that it reproduces the energy release following $t^{-1.2}$.
Here $t_{\rm 0, decay}$ is the beginning of the radioactive energy release.
In this paper, we set $t_{\rm 0, decay} = 10^{-4}$ days,
which is sufficiently earlier than the initial time of the simulations.
When the computation reaches to $t > t_{\rm active}$, 
UVOIR packets are created 
(instead of $\gamma$-ray packet for the case of SNe,
see Section \ref{sec:models}).
Similar to the case of SNe, 
an isotropic direction in comoving frame is assigned.
For the UVOIR packets, the initial co-moving wavelength 
is assigned by sampling emissivity
$j_{\lambda}$ (Section \ref{sec:events}).

Both for the cases of SNe and NS mergers, 
when the time of the activation of a packet 
is earlier than the initial time of the simulation 
($t_{\rm active} < t_0$, where $t_0$ is the initial time of the simulation), 
the packet is created as a UVOIR packet at $t=t_0$.
To take into account the energy loss by adiabatic expansion, 
the comoving-frame energy $\epsilon$ is reduced to 
$\epsilon = \epsilon_0 (t_{\rm active}/t_0)$ \citep{lucy05}.

Note that the current code does not take into account 
the heating by the shock wave (see \eg \citealt{kasen06}).
Thus, the code cannot be applied for Type IIP SNe, where 
the shock heating is a dominant source of radiation 
at the plateau phase (up to $\sim 100$ days).

\subsection{$\gamma$-ray transfer}
\label{sec:gamma}

For the case of SNe, $\gamma$-ray transfer is computed.
We adopt the gray approximation with a mass absorption coefficient
of $\kappa_{\gamma} = 0.027\ {\rm cm^2 \ g^{-1}}$,
which is known to reproduce the results of multi-energy transport
and the observed light curves of Type Ia SNe
\citep{colgate80,sutherland84,maeda06gamma}.
This is also confirmed by our test calculations (Appendix A).
Once a $\gamma$-ray packet is absorbed, 
it is immediately converted to a UVOIR packet.
For the case of NS mergers, the effect of $\gamma$-ray transport
is taken into account by introducing a thermalization factor, 
and $\gamma$-ray transfer is not directly computed
(see Section \ref{sec:models}).

\subsection{UVOIR transfer}
\label{sec:UVOIR}

Transfer of UVOIR packets is computed 
considering a wavelength-dependent opacity.
As opacity sources, we consider
the electron scattering, and free-free, bound-free, and bound-bound transitions.
The wavelength-dependent opacity is evaluated in each cell 
after the temperature estimate in each time step.
The bound-bound transition is the dominant source of opacity
both for Type Ia SNe and NS mergers.

{\bf Electron scattering: }
By solving the Saha equations, the number density of free electrons ($n_e$)
is computed in each cell.
The absorption coefficient of electron scattering is evaluated 
as $\alpha^{\rm es} = n_e \sigma_{\rm Th}$, where $\sigma_{\rm Th}$ 
is the cross section of electron scattering (or Thomson scattering).

{\bf Free-free transition: }
Free-free absorption coefficient 
for an ion ($i$-th element and $j$-th ionization stage) 
is computed as in \citet{rybicki79}, using common convention;
\begin{eqnarray}
\alpha_{i,j}^{\rm ff} (\lambda) &=& 
\frac{4e^6}{3m_e hc} \left( \frac{2 \pi}{3 k m_e}\right)^{1/2}
T^{-1/2} (j-1)^2 \nonumber \\
&& n_e n_{i,j} \nu^{-3} (1 - e^{-h\nu/kT}) \bar{g}_{\rm ff},
\end{eqnarray}
where $T$, $n_{i,j}$ are the electron temperature 
(which is assumed to be the same with radiation temperature, 
Section \ref{sec:temperature}), and the number density of the ion.
Here $\bar{g}_{\rm ff}$ is a velocity-averaged Gaunt factor,
which is set to be unity in our code.
The absorption coefficient is evaluated for all the ions 
included in the ejecta.

{\bf Bound-free transition: }
For the bound-free absorption coefficients,
we adopt 
\begin{equation}
\alpha_{i,j}^{\rm bf} (\lambda) = n_{i,j} \sigma_{i,j}^{\rm bf},
\end{equation}
where $\sigma_{i,j}^{\rm bf}$ is the cross section of
bound-free transition for an ion.
For the cross section,
we use analytic formulae by \citet{verner96},
which is expressed by 7 parameters.
They cover elements from H through Si, and S, Ar, Ca and Fe 
for all the ionization stages.
The missing data are replaced with those of the closest elements.

Since there is no data for elements heavier than Fe,
we simply use the cross section of Fe for all the heavier elements.
This crude assumption does not affect our conclusions
for the emission from NS mergers. 
The bound-free opacity can be dominant only at 
$\lambda \lsim 1000$ \AA, 
while a typical radiation temperature we treat is $T_R < 10,000$ K.
Thus, the bound-free opacity does not have a strong impact
on the overall properties of the radiation.

{\bf Bound-bound transition: }
The bound-bound transition is treated using 
``the expansion opacity'' introduced by \citet{karp77}.
We adopt the formula by \citet{eastman93};
\begin{equation}
\alpha_{\rm exp}^{bb} (\lambda) = \frac{1}{ct} 
\sum_l \frac{\lambda_l}{\Delta \lambda} (1 - e^{- \tau_l}),
\end{equation}
which is also adopted by \citet{kasen06}.
The summation is taken for all the lines within
a wavelength interval of $\Delta \lambda$.
Here, $\lambda_l$ and $\tau_l$ are the wavelength and 
the Sobolev optical depth of a transition, respectively.
In homologously expanding ejecta, 
the Sobolev optical depth can be written as 
\begin{equation}
\tau_l = \frac{\pi e^2}{m_e c} 
\left( \frac{n_{i,j} \lambda_l t}{g_0} \right) g_l f_l \exp^{-E_l/kT}.
\end{equation}
Here $g_l$, $E_l$, $f_l$ are the statistical weight 
and the energy of the lower level of the transition, 
and the oscillator strength of the transition, respectively
($g_0$ is the statistical weight for the ground level).

For the properties of bound-bound transitions 
($\lambda_l$, $g_l$, $E_l$, and $f_l$),
the line list by \citet[CD23]{kurucz95} is adopted.
This list includes about 500,000 lines,
and has been widely used for radiative transfer of SNe 
\citep{kasen06,kromer09}. 
In Appendix A, we demonstrate that this list gives 
reasonable light curves and spectra of Type Ia SNe.

For NS mergers, the dominant elements in the ejecta 
may be r-process elements (with $Z \ge 31$).
For such heavy elements, Kurucz's line list includes the data 
only for neutral or some singly ionized ions.
Thus, we constructed a line list for elements heavier than 
Ga ($Z=31$) using the VALD database 
\citep{piskunov95,ryabchikova97,kupka99,kupka00}
\footnote{\url{http://vald.astro.univie.ac.at/~vald/php/vald.php}}.
Figure \ref{fig:nline} shows the number of bound-bound transitions
as a function of atomic number.
Our list includes the data up to doubly ionized ions.
In total, about 100,000 lines are added at $Z \ge 31$.
The limitation of this line list 
is discussed in Appendix B.

We treat all the bound-bound transition as purely absorptive
(see discussion by \citealt{nugent97}).
As demonstrated by \citet{kasen06}, this choice gives a reasonable 
spectral series and light curves for Type Ia SNe.
We also confirmed this in our test calculations (see Appendix A).

\subsection{Treatment of events}
\label{sec:events}
In MC radiative transfer, propagation of packets is directly followed.
To evaluate what kind of events packets experience,
3 distances are computed for each packet:
(1) the distance to the scattering or absorption events $l_1$,
(2) the distance to the next cell $l_2$, and
(3) the distance that a packet can travel before the next time step
$l_3$.
For the distance to the scattering or absorption events,
we first set a threshold optical depth $\tau_{\rm th} = -\ln z$.
Then, by using the total absorption coefficient for UVOIR photons
\begin{equation}
\alpha^{\rm tot} (\lambda) = \alpha^{\rm es} +
\sum_{i,j} \alpha_{i,j}^{\rm ff} (\lambda) +
\sum_{i,j} \alpha_{i,j}^{\rm bf} (\lambda) +
\alpha_{\rm exp}^{\rm bb} (\lambda), 
\end{equation}
the distance is computed from 
$l_1 = \tau_{\rm th}/\alpha^{\rm tot} (\lambda)$,
so that it reproduces the attenuation following $\exp (-\tau)$.
For the $\gamma$-ray transfer, 
$\alpha^{\rm tot} = \kappa_{\rm \gamma} \rho$.

Among the three distances, 
the event with the shortest distance occurs.
When $l_1$ is the shortest, we judge if it is a scattering event 
or an absorption event.
For the $\gamma$-ray transfer, it is always an absorption event.
For the UVOIR transfer, only the electron scattering is a scattering event,
so that an event is treated as scattering 
if $z < \alpha^{\rm es}/\alpha^{\rm tot}$.
For the scattering event, 
the co-moving wavelength and energy of a packet are conserved.
For the absorption event, 
the next co-moving wavelength is determined by sampling the emissivity
(by Kirchhoff's law)
\begin{equation}
j_{\lambda} = \alpha^{\rm abs} (\lambda) B_{\lambda} (T),
\end{equation}
where $\alpha^{\rm abs}$ is the total coefficient for absorptive events
\begin{equation}
\alpha^{\rm abs} (\lambda) = 
\sum_{i,j} \alpha_{i,j}^{\rm ff} (\lambda) +
\sum_{i,j} \alpha_{i,j}^{\rm bf} (\lambda) +
\alpha_{\rm exp}^{\rm bb} (\lambda).
\end{equation}

When $l_2$ is the shortest, the comparison among $l_1, l_2$, and $l_3$ 
is repeated again in the next cell.
These procedures are repeated until $l_3$ becomes the shortest.
When $l_3$ is the shortest, 
the position and direction of the packet are recorded, 
and the propagation of the next packet is considered.
After computing the propagation of all the packets in a time step,
the propagation in the next time step is computed.

\subsection{Temperature determination}
\label{sec:temperature}

After the propagation of the packets, 
the temperature in each cell is evaluated by using the photon flux.
In MC transfer, the photon intensity is evaluated 
\citep{lucy03} as 
\begin{equation}
J_{\nu} d\nu = \frac{1}{4 \pi \Delta t V}\sum_{d\nu} \epsilon ds.
\end{equation}
The temperature is estimated by approximating that
the wavelength-integrated intensity $<J> = \int J_{\nu} d\nu$ 
follows Stefan-Boltzmann law, \ie
\begin{equation}
<J> = \frac{\sigma}{\pi}T_{R}^4.
\end{equation}
This is the same assumption with the "simple" case of \citet{kromer09}.
It is confirmed that this method gives reasonable results in Appendix A.
We assume that the kinetic temperature of electron $T_e$ is 
the same with the radiation temperature $T_R$.
We simply denote them by $T$, \ie $ T = T_e = T_R$.

\subsection{Ionization and excitation}
\label{sec:ionization}

For ionization, we assume LTE and 
solve the Saha equations for H through U simultaneously.
We use NIST database 
\footnote{\url{http://www.nist.gov/pml/data/asd.cfm}}
for the atomic data, such as partition functions
and ionization potentials.
For the excitation, we assume Boltzmann distribution with 
the temperature $T$.

In the NS merger ejecta, 
radioactive decay produces fast $\beta$-decay electrons, 
fission products, and gamma-rays in the NS merger ejecta.
Thus, possible deviation from LTE by these non-thermal ionization
and excitation processes may be expected.
In fact, non-thermal effect is known to be important 
for excitation of He lines in Type Ib SNe 
\citep{lucy91,dessart11,hachinger12}.
However, \citet{kasen13} estimated that 
the non-thermal excitation rate in a typical environment 
of the NS merger ejecta (blackbody temperature of 5000 K at $t =$ 1 day)
is only $\sim 10^{-8}$ of the radiative excitation rate by blackbody field.
This implies that non-thermal processes do not 
affect ionization and excitation states significantly.
Nevertheless, it must be noted that the deviation from LTE is expected 
to be larger at later epochs as the blackbody temperature decreases 
and the ejecta becomes more transparent.

\subsection{Observations}
\label{sec:observations}

Escaping $\gamma$-ray packets and UVOIR packets are counted.
They naturally give a $\gamma$-ray light curve and 
a UVOIR light curve.
Since each UVOIR packet has a wavelength, 
time series of spectra are also obtained.
Multi-color light curves are computed 
by adopting the standard optical $UBVRI$ filters 
\citep{bessell90} and NIR $JHK$ filters \citep{persson98}
with the zero-magnitude flux by \citet{bessell98}.
AB magnitudes using Sloan Digital Sky Survey $ugriz$ filters
\citep{fukugita96} and $JHK$ filters are also computed.

%%%%%%%%%%%%%%%%%%%%%%%%%%%%%%%%%%%%%%%%%%%%%%%%%%%% 
% Table: Model    %%%%%%%%%%%%%%%%%%%%%%%%%%%%%%%%%% 
%%%%%%%%%%%%%%%%%%%%%%%%%%%%%%%%%%%%%%%%%%%%%%%%%%%% 
\begin{deluxetable*}{ccccc} 
%\rotate
\tablewidth{0pt}
\tablecaption{Summary of Models}
\tablehead{
Model & $\Mej$    & $\KE$     & $v_{\rm ch}$ & Abundance$^{1}$ \\
      & ($\Msun$) & (erg)     &             &           
}
\startdata
NSM-all   & $1.0 \times 10^{-2}$  & $1.3 \times 10^{50}$  &  0.12$c$  & $31 \le Z \le 92$ \\
NSM-dynamical & $1.0 \times 10^{-2}$  & $1.3 \times 10^{50}$  &  0.12$c$  & $55 \le Z \le 92$ \\
NSM-wind  & $1.0 \times 10^{-2}$  & $1.3 \times 10^{50}$  &  0.12$c$  & $31 \le Z \le 54$ \\
NSM-Fe    & $1.0 \times 10^{-2}$  & $1.3 \times 10^{50}$  &  0.12$c$  & $Z=26$ (only Fe) \\ \hline

APR4-1215  &  $8.6 \times 10^{-3}$    &  $4.3 \times 10^{50}$  & 0.24$c$  & $31 \le Z \le 92$ \\
APR4-1314  &  $8.1 \times 10^{-3}$    &  $3.6 \times 10^{50}$  & 0.22$c$  & $31 \le Z \le 92$ \\
H4-1215  &  $3.5 \times 10^{-3}$      &  $1.4 \times 10^{50}$  & 0.21$c$  & $31 \le Z \le 92$ \\
H4-1314  &  $7.0 \times 10^{-4}$      &  $1.9 \times 10^{49}$  & 0.17$c$  & $31 \le Z \le 92$ \\
\enddata
\tablecomments{
$^1$ Solar abundance ratios \citep{simmerer04} are assumed.
}
\label{tab:models}
\end{deluxetable*}
%%%%%%%%%%%%%%%%%%%%%%%%%%%%%%%%%%%%%%%%%%%%%%%%%%%% 

\section{Models and Setup}
\label{sec:models}

\subsection{Ejecta Models}
\label{sec:models1}

We adopt two kinds of models for NS merger ejecta 
(Table \ref{tab:models}).
The first sets are simple models with a power-law density structure,
and the others are realistic models based on the results of 
merger simulations by \citet{hotokezaka13}.

For simple cases, we use a model
introduced by \citet{metzger10}.
The density structure follows $\rho \propto r^{-3}$
from $v=0.05 c$ - $0.2c$.
The total ejecta mass is set to be $\Mej = 0.01 \Msun$.
This model has a total kinetic energy of $\KE = 1.3 \times 10^{50}$ erg.
We define the characteristic velocity of the ejecta by
$v_{\rm ch} = \sqrt{2\KE/\Mej}$.
Then, the characteristic velocity of this model is $v_{\rm ch} = 0.12c$.
The ejecta include a mixture of r-process elements, \ie Ga through U.
For relative mass fractions, we assume the solar abundance ratios
of r-process elements by \citet{simmerer04}.
Hereafter, we call this fiducial model ``NSM-all''

To see the effect of element abundances to the opacity,
we also test additional three models:
(1) ``NSM-dynamical'': a model only with $Z \ge 55$,
which may be realized when the r-process nucleosynthesis is efficient
in the dynamical ejecta, 
and the final element abundances are dominantly determined by 
fission cycles.
(2) ``NSM-wind'': a model with $31 \le Z \le 54$.
These relatively light elements can be abundant
if the wind or outflow from a black hole torus,
which can subsequently occur after the NS merger, dominates the ejecta 
\citep{wanajo12,fernandez13}.
(3) ``NSM-Fe'': a hypothetical model only with Fe.
To see the effect to the opacity, the heating rate is kept the same
in these models (Section \ref{sec:models2}).

The other sets of models are 
constructed from results of numerical simulations.
\citet{hotokezaka13} performed extensive numerical-relativity 
simulations for various mass of neutron stars and 
EOSs.
We map the density distribution of the ejecta (the material that
has a higher expansion velocity than the escape velocity
at the end of the simulations) into
a two-dimensional, axisymmetric model.
We adopt 4 models from their simulations (Table \ref{tab:models}).
The adopted EOSs are a ``soft'' EOS APR4 \citep{akmal98}
and a ``stiff'' EOS H4 \citep{glendenning91,lackey06},
which give the radius of 11.1 km and 13.6 km
for a $1.35 \Msun$ neutron star (in gravitational mass), respectively.
In this paper, we use terminology of ``soft'' for 
EOSs giving smaller radii of NSs.
The gravitational masses of NSs are
$1.2 \Msun\ + 1.5 \Msun$ and  $1.3 \Msun + 1.4 \Msun$.
\footnote{In this paper, the mass ratio is defined as $M_1/M_2$, 
where $M_1$ and $M_2$ are the mass of two NSs and $M_1 > M_2$.}
See \citet{hotokezaka13} for more details.
\footnote{The kinetic energy shown in Table \ref{tab:models}
is somewhat different from that shown by \citet{hotokezaka13}
because we smoothed the density structure of numerical results
and remapped it into coarse grids.}

\subsection{Setup}
\label{sec:models2}

The major difference between SN ejecta and NS merger ejecta 
comes from the heating source.
SN ejecta are dominantly heated by the decay of \Nifs\ while 
NS merger ejecta are heated by the decay of many different 
radioactive r-process nuclei with different decay timescales.
\citet{metzger10} showed that the energy release per unit mass
is $\sim 3 \times 10^{15} (t/t_{\rm 0, decay})^{-1.2}\ {\rm erg\ s^{-1}\ g^{-1}}$,
which does not depend strongly on the models
(see also \citealt{korobkin12}).
Thus, we assume that the total radioactive power is proportional 
to the total ejecta mass, and adopt
\begin{equation}
\dot{E}_{\rm decay} = 6 \times 10^{46} 
\left( \frac{\Mej}{0.01 \Msun} \right) 
\left( \frac{t}{t_{\rm 0, decay}} \right)^{-1.2}\ {\rm erg\ s^{-1}}.
\label{eq:heating}
\end{equation}
As introduced in Section \ref{sec:packets}, 
we set $t_{\rm 0, decay} = 10^{-4}$ days.
As long as it is sufficiently earlier than the initial time of the 
simulations, this choice does not affect the light curve
since most of the energy released at such an early stage
is lost by adiabatic expansion.

From this total power, 
about $90 \%$ of the energy is released by $\beta$ decay
while the other $10 \%$ by fission \citep{metzger10}.
For the $\beta$ decay, 
neutrinos, electrons, and $\gamma$-rays carry 
about $25 \%$, $25 \%$, and $50 \%$ 
of the energy, respectively.
The energy carried out by electrons and by fission products 
is likely to be deposited without significant escape, 
while that by neutrinos can escape almost entirely.
Following the method by \citet{metzger10} and \citet{korobkin12},
we introduce the efficiency of thermalization
$\epsilon_{\rm therm}$ for the $\beta$-decay energy.
Then, the radiation energy can be written as 
\begin{equation}
\dot{E}_{\rm rad} = (0.1 + 0.9 \epsilon_{\rm therm}) \dot{E}_{\rm decay}.
\end{equation}
The thermalization efficiency must be in the range of 
$\epsilon_{\rm therm} = 0.25$ - 0.75, depending on the 
efficiency of $\gamma$-ray energy deposition.
By adopting $\epsilon_{\rm therm} = 0.5$,
we assume that the energy $\dot{E}_{\rm rad}$
is immediately deposited without transportation
(\ie we neglect $\gamma$-ray transfer, 
and UVOIR packets are created directly).

%%%%%%%%%%%%%%%%%%%%%%%%%%%%%%%%%%%%%%%%%%%%%%%%%%%% 
% Figure  %%%%%%%%%%%%%%%%%%%%%%%%%%%%%%%%%%%%%%%%%% 
%%%%%%%%%%%%%%%%%%%%%%%%%%%%%%%%%%%%%%%%%%%%%%%%%%%% 
\begin{figure}
\begin{center}
\includegraphics[scale=1.4]{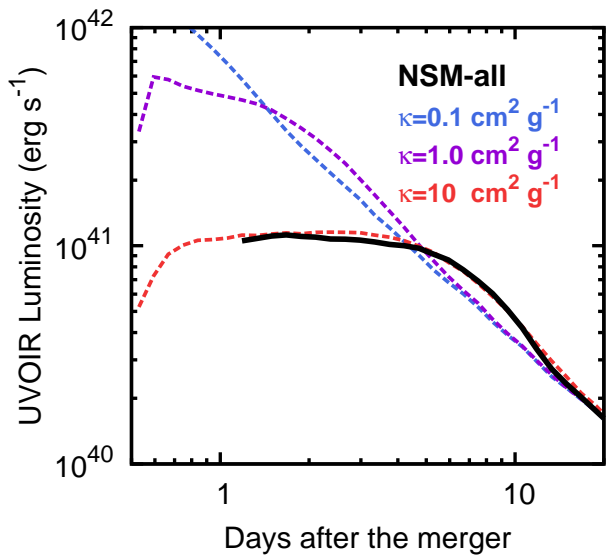} 
\caption{
Bolometric light curve of model NSM-all
(black, multi-frequency simulations).
It is compared with the light curves for the same model
but with the gray approximation of UVOIR transfer 
($\kappa$ =0.1, 1, and 10 ${\rm cm^2\ g^{-1}}$ 
for blue, purple, and red lines, respectively).
The result of multi-frequency transfer are most similar to 
that of gray transfer with $\kappa$ = 10 ${\rm cm^2\ g^{-1}}$.
\label{fig:NSM-all}}
\end{center}
\end{figure}
%%%%%%%%%%%%%%%%%%%%%%%%%%%%%%%%%%%%%%%%%%%%%%%%%%%% 

%%%%%%%%%%%%%%%%%%%%%%%%%%%%%%%%%%%%%%%%%%%%%%%%%%%% 
% Figure  %%%%%%%%%%%%%%%%%%%%%%%%%%%%%%%%%%%%%%%%%% 
%%%%%%%%%%%%%%%%%%%%%%%%%%%%%%%%%%%%%%%%%%%%%%%%%%%% 
\begin{figure*}
\begin{center}
\includegraphics[scale=1.4]{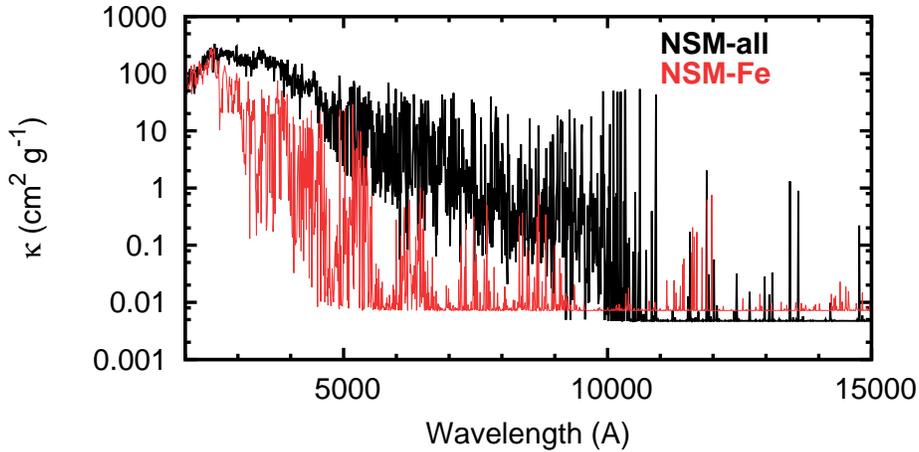} 
\caption{
Mass absorption coefficient $\kappa$ 
at $v=0.1 c$ in model NSM-all and NSM-Fe 
as a function of wavelength ($t=3$ days after the merger).
In r-process element-rich ejecta, the opacity is 
higher than Fe-rich ejecta by factor of about 100
around the center of optical wavelength ($\sim 5000$ \AA).
\label{fig:kappa}}
\end{center}
\end{figure*}
%%%%%%%%%%%%%%%%%%%%%%%%%%%%%%%%%%%%%%%%%%%%%%%%%%%% 

\section{Results}
\label{sec:simple}

Figure \ref{fig:NSM-all} shows the computed bolometric light curve
for the fiducial model NSM-all (black line).
It is compared with the light curves 
for the same model but with the gray approximation of the UVOIR transfer.
The blue, purple, and red lines show the cases with 
gray mass absorption coefficients 
of $\kappa = 0.1$, 1.0, and 10 ${\rm cm^2\ g^{-1}}$, respectively.
The result of multi-frequency transfer closely follows the 
light curve with the gray opacity of $\kappa=10\ {\rm cm^2\ g^{-1}}$.
This indicates that r-process element-rich NS merger ejecta 
are more opaque than previously assumed 
($\kappa \simeq 0.1 \ {\rm cm^2\ g^{-1}}$,
\eg \citealt{li98,metzger10}), by a factor of about 100.
As a result, the bolometric light curve becomes fainter, 
and the time scale becomes longer.
\footnote{We show the results of our multi-frequency transfer simulations 
at $t \gsim 1$ day. Because of the lack of bound-bound transition data 
for triply ionized ions in our line list (Figure \ref{fig:nline}), 
the opacity at earlier epoch is not correctly evaluated. 
We hereafter show the results when the temperature 
at the characteristic velocity is below 10,000 K, when
the dominant ionization states are no more triply ionized ions.
Detailed discussion is presented in Appendix B.}
This is consistent with the findings by \citet{kasen13} 
and \citet{barnes13}.

Figure \ref{fig:kappa} shows the mass absorption coefficient 
as a function of wavelength at $t=3$ days in model NSM-all at $v=0.1c$.
The mass absorption coefficient is 
as high as 1-100 ${\rm cm^2\ g^{-1}}$ in the optical wavelengths.
The resulting Planck mean mass absorption coefficient is 
about $\kappa = 10\ {\rm cm^2\ g^{-1}}$ (Figure \ref{fig:NSM_early}).
This is the reason why the bolometric light curve
of multi-frequency transfer most closely follows that 
with gray opacity of $\kappa = 10\ {\rm cm^2\ g^{-1}}$ in Figure \ref{fig:NSM-all}.

\begin{figure}
\begin{center}
\includegraphics[scale=1.4]{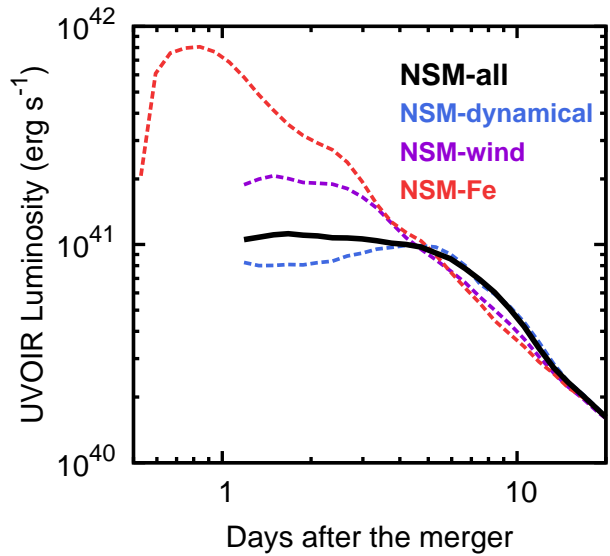} 
\caption{
Bolometric light curves for 
simple models with different element abundances:
NSM-all ($31 \le Z \le 92$), NSM-dynamical ($55 \le Z \le 92$),
NSM-wind ($31 \le Z \le 54$), and NSM-Fe (only Fe).
\label{fig:simple}}
\end{center}
\end{figure}

The high opacity in r-process element-rich ejecta is also 
confirmed by the comparison with other simple models.
Figure \ref{fig:simple} shows the comparison of the 
bolometric light curve
among models NSM-all, NSM-dynamical, NSM-wind, and NSM-Fe.
Compared with NSM-Fe, the other models show the fainter light curves.
This indicates that the elements heavier than Fe 
contribute to the high opacity.
The opacity in model NSM-Fe is also shown in Figure \ref{fig:kappa}.
It is nicely shown that opacity in NSM-all is 
higher than that in NSM-Fe by a factor of about 100 
at the center of optical wavelengths ($\sim 5000$ \AA).

As inferred from Figure \ref{fig:simple}, 
NSM-dynamical ($55 \le Z \le 92$) has a higher opacity 
than that of NSM-wind ($31 \le Z \le 54$).
This is because lanthanoid elements ($57 \le Z \le 71$) have 
the largest contribution to the bound-bound opacity,
as demonstrated by \citet{kasen13}.
Note that, however, 
even with the elements at $31 \le Z \le 54$,
the opacity is higher than that of Fe.

%%%%%%%%%%%%%%%%%%%%%%%%%%%%%%%%%%%%%%%%%%%%%%%%%%%% 
% Figure  %%%%%%%%%%%%%%%%%%%%%%%%%%%%%%%%%%%%%%%%%% 
%%%%%%%%%%%%%%%%%%%%%%%%%%%%%%%%%%%%%%%%%%%%%%%%%%%% 
%%%%%%%%%%%%%%%%%%%%%%%%%%%%%%%%%%%%%%%%%%%%%%%%%%%% 

%%%%%%%%%%%%%%%%%%%%%%%%%%%%%%%%%%%%%%%%%%%%%%%%%%%% 
% Figure  %%%%%%%%%%%%%%%%%%%%%%%%%%%%%%%%%%%%%%%%%% 
%%%%%%%%%%%%%%%%%%%%%%%%%%%%%%%%%%%%%%%%%%%%%%%%%%%% 
\begin{figure}
\begin{center}
\includegraphics[scale=1.4]{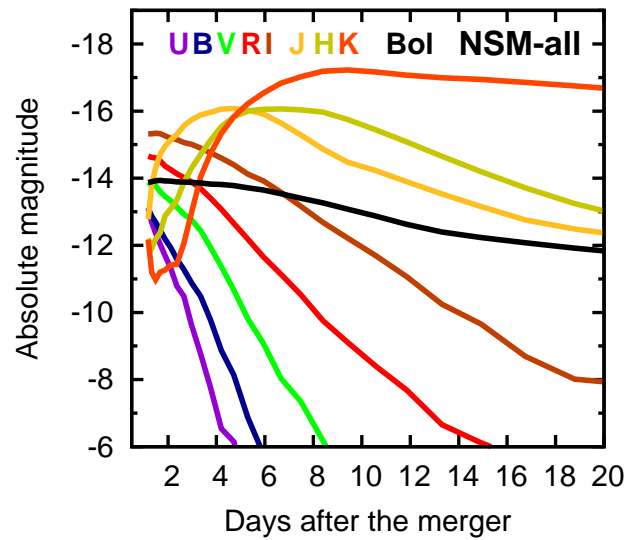} 
\caption{
Multi-color light curves of model NSM-all (in Vega magnitude).
Light curves in redder bands are brighter and slower.
\label{fig:multicolor}}
\end{center}
\end{figure}
%%%%%%%%%%%%%%%%%%%%%%%%%%%%%%%%%%%%%%%%%%%%%%%%%%%% 

%%%%%%%%%%%%%%%%%%%%%%%%%%%%%%%%%%%%%%%%%%%%%%%%%%%% 
% Figure  %%%%%%%%%%%%%%%%%%%%%%%%%%%%%%%%%%%%%%%%%% 
%%%%%%%%%%%%%%%%%%%%%%%%%%%%%%%%%%%%%%%%%%%%%%%%%%%% 
\begin{figure*}
\begin{center}
\includegraphics[scale=1.4]{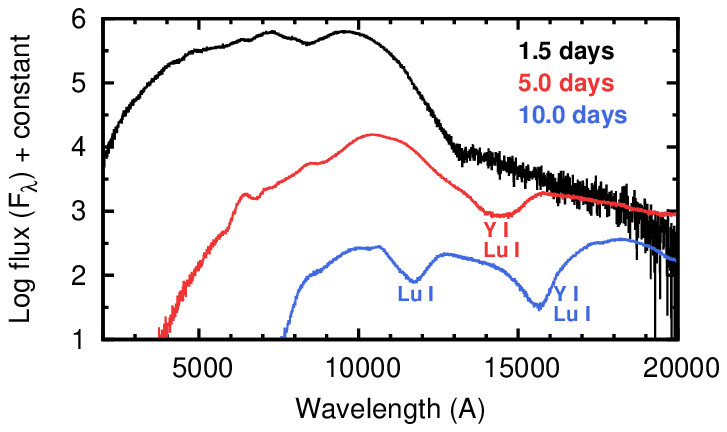} 
\caption{
UVOIR spectra of model NSM-all at $t$ = 1.5, 5.0, and 10.0 days
after the merger.
The spectra are almost featureless.
At NIR wavelengths, there are possible absorption troughs,
which result from \ion{Y}{i}, \ion{Y}{ii}, and \ion{Lu}{i} in our simulations.
However, the features could be result from 
the incompleteness of the line list in the NIR wavelengths.
\label{fig:spec}}
\end{center}
\end{figure*}
%%%%%%%%%%%%%%%%%%%%%%%%%%%%%%%%%%%%%%%%%%%%%%%%%%%% 

Figure \ref{fig:multicolor} shows the multi-color light curves
of model NSM-all.
In general, the emission from NS merger ejecta is red
because of (1) a lower temperature than SNe and 
(2) a higher optical opacity than in SNe.
In particular, the optical light curves in the blue wavelengths
drop dramatically in the first 5 days.
The light curves in the redder band evolves more slowly.
This trend is also consistent with the results by \citet{kasen13}
and \citet{barnes13}.

Since our simulations include all the r-process elements, 
spectral features are of interest.
Since the simulations by \citet{kasen13} and \citet{barnes13} include
only a few lanthanoid elements, 
they do not discuss the detailed spectral features.
Figure \ref{fig:spec} shows the spectra of model NSM-all at 
$t = 1.5$, 5.0 and 10.0 days after the merger.
Our spectra are almost featureless at all the epochs.
This is because of the overlap of many bound-bound transitions of 
different r-process elements.
As a result, compared with the results by
\citet{kasen13} and \citet{barnes13}, 
the spectral features are more smeared out.

Note that we could identify possible broad absorption features
around 1.4 $\mu$m (in the spectrum at $t=5$ days)
and around 1.2 $\mu$m and 1.5 $\mu$m ($t=10$ days).
In our line list, these bumps are mostly made by a cluster of 
the transitions of \ion{Y}{I}, \ion{Y}{II}, and \ion{Lu}{I}.
However, we are cautious about such identifications because
the bound-bound transitions in the VALD database are not likely 
to be complete in the NIR wavelengths even for neutral and singly ionized
ions.
In fact, \citet{kasen13} showed that the opacity of Ce from the 
VALD database drops in the NIR wavelengths,
compared with the opacity based on their atomic models.
Although we cannot exclude a possibility that 
a cluster of bound-bound transitions of some ions
can make a clear absorption line in NS mergers,
our current simulations do not provide prediction for such features.

%%%%%%%%%%%%%%%%%%%%%%%%%%%%%%%%%%%%%%%%%%%%%%%%%%%% 
% Figure  %%%%%%%%%%%%%%%%%%%%%%%%%%%%%%%%%%%%%%%%%% 
%%%%%%%%%%%%%%%%%%%%%%%%%%%%%%%%%%%%%%%%%%%%%%%%%%%% 
\begin{figure}
\begin{center}
\includegraphics[scale=1.4]{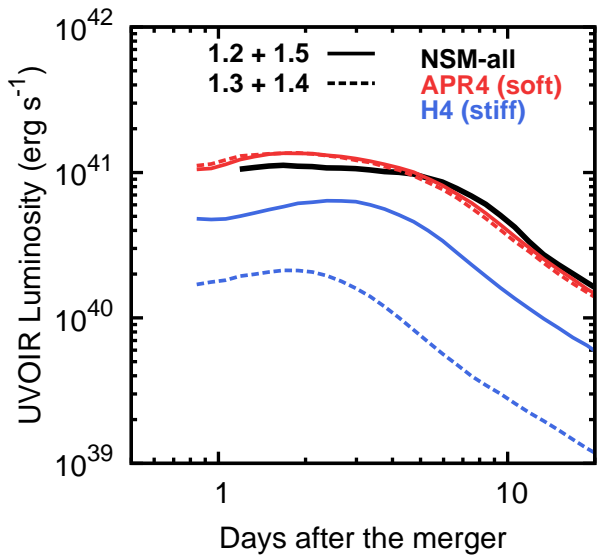} 
\caption{
Bolometric light curves for realistic models (Table \ref{tab:models}).
The luminosity is averaged over all solid angles.
The expected emission of models with a soft EOS APR4 (red)
is brighter than that with a stiff EOS H4 (blue).
For the soft EOS APR4, the light curve does not depend on the
mass ratio, while for a stiff EOS H4, a higher mass ratio 
($1.2 \Msun\ + 1.5 \Msun$, solid line)
results in a large ejecta mass, and thus, brighter emission
than a lower mass ratio ($1.3 \Msun\ + 1.4 \Msun$, dashed line).
\label{fig:realistic}}
\end{center}
\end{figure}
%%%%%%%%%%%%%%%%%%%%%%%%%%%%%%%%%%%%%%%%%%%%%%%%%%%% 

\section{Dependence on the EOS and Mass Ratio}
\label{sec:realistic}

Figure \ref{fig:realistic} shows the bolometric light curves 
of realistic models.
The luminosity is averaged over all solid angles.
Since the angle dependence is within a factor of 2 
(brighter in the polar direction, see \citealt{roberts11}), 
we focus only on the averaged luminosity.

The models with the soft EOS APR4 (red) is brighter
than the models with the stiff EOS H4 (blue).
This is interpreted as follows.
When the total radioactive power is propotional to the ejecta mass
(Equation \ref{eq:heating}), the peak luminosity is expected to scale 
as $L \propto \Mej^{1/2} v_{\rm ch}^{1/2}$ \citep{li98}.
We confirmed that the peak luminosity of our models
roughly follows this relation (an effective opacity is 
$\kappa \sim 10 \ {\rm cm^2 g^{-1}}$ irrespective of models).
For a soft EOS (\ie a smaller radius of a NS), 
the mass ejection occurs at a more compact orbit
and shock heating is efficient.
As a result, the mass of the ejecta is higher for softer EOSs 
\citep[see Table \ref{tab:models}, and also][]{hotokezaka13,bauswein13}.
Therefore, the NS merger with the soft EOS APR4 is brighter.
Note that the light curve of the fiducial model NSM-all (black) 
is similar to that of model APR4-1215 and APR4-1314
because these models have a similar mass 
and a characteristic velocity (Table \ref{tab:models}).

For the soft EOS APR4, the brightness does not 
depend strongly on the mass ratio of the binary NSs 
(red solid and dashed lines in Figure \ref{fig:realistic}).
This is because for a soft EOS, such as APR4, 
the mass ejection by shock heating is efficient.
By contrast, for the stiff EOS H4, 
the mass ejection occurs primarily by tidal effects 
(the effect of shock heating is weak, \citealt{hotokezaka13}).
Thus, the mass ejection is more efficient for a higher mass ratio.
As a result, model H4-1215 (mass ratio of 1.25)
is brighter than model H4-1314 (mass ratio of 1.08).

These results open a new window to study 
the nature of the NS merger and EOSs.
By adding the information of EM radiation
to the analysis of GW signals,
we may be able to pin down the masses of two NSs 
and/or stiffness of the EOSs more accurately.
Note that, in the current simulations, 
the heating rate per mass is fixed.
To fully understand the connection between the 
initial condition of the NS merger and expected emission,
detailed nucleosynthesis calculations are necessary.

%%%%%%%%%%%%%%%%%%%%%%%%%%%%%%%%%%%%%%%%%%%%%%%%%%%% 
% Figure  %%%%%%%%%%%%%%%%%%%%%%%%%%%%%%%%%%%%%%%%%% 
%%%%%%%%%%%%%%%%%%%%%%%%%%%%%%%%%%%%%%%%%%%%%%%%%%%% 
\begin{figure*}
\begin{center}
\begin{tabular}{cc}
\includegraphics[scale=1.05]{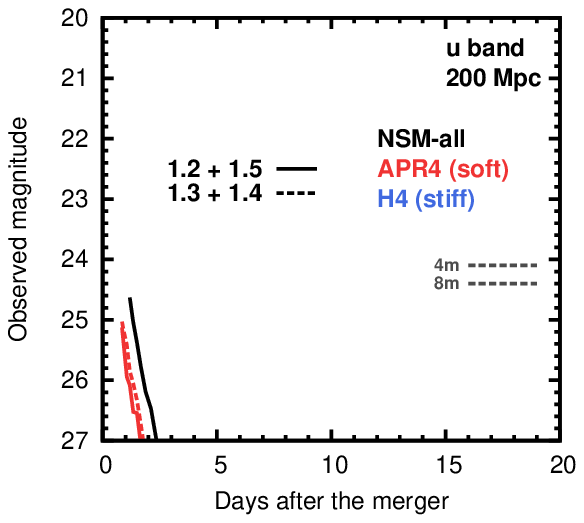} &
\includegraphics[scale=1.05]{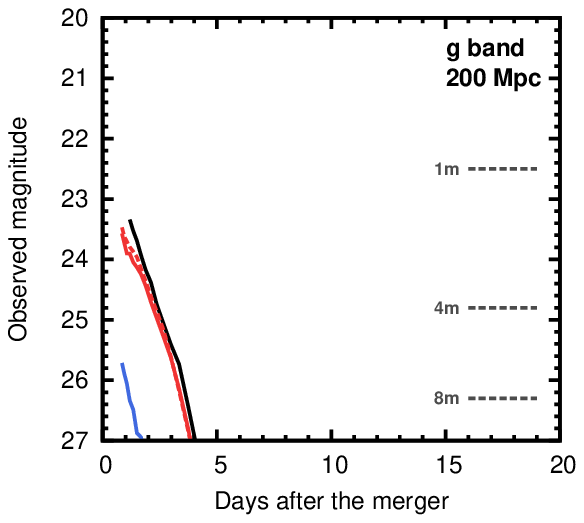} \\
\includegraphics[scale=1.05]{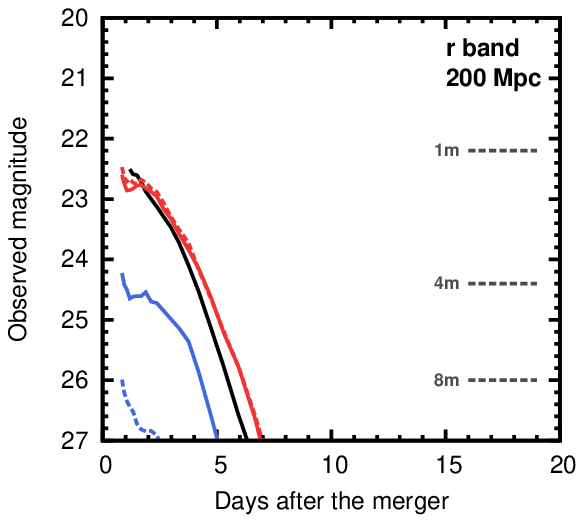} &
\includegraphics[scale=1.05]{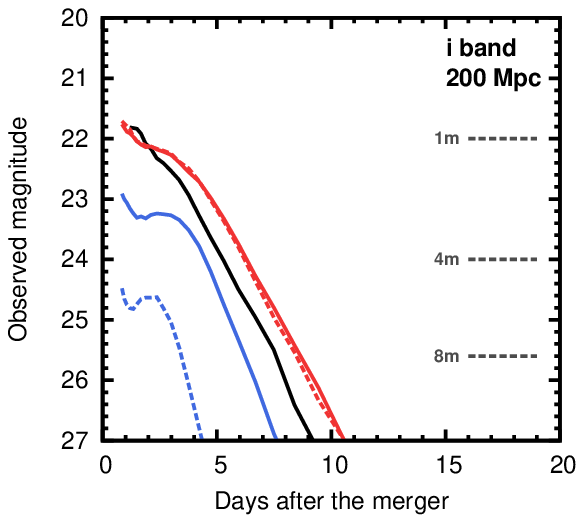} \\
\includegraphics[scale=1.05]{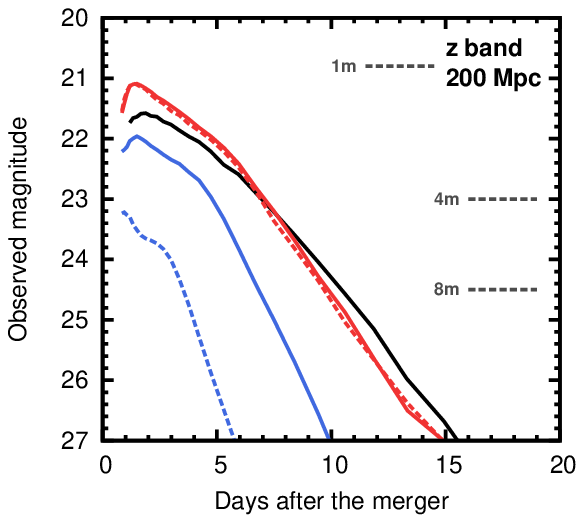} & 
\includegraphics[scale=1.05]{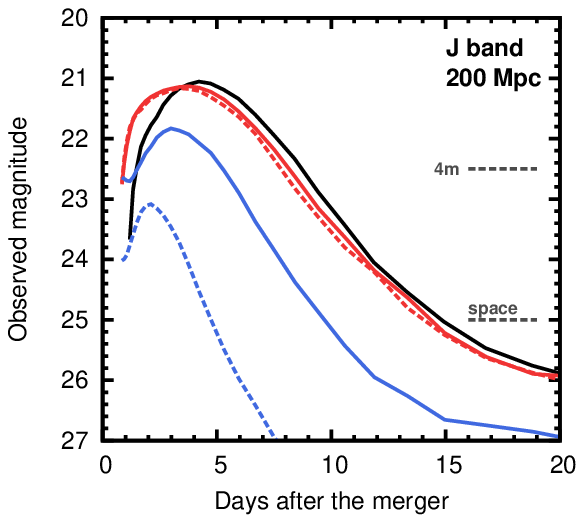} \\
\includegraphics[scale=1.05]{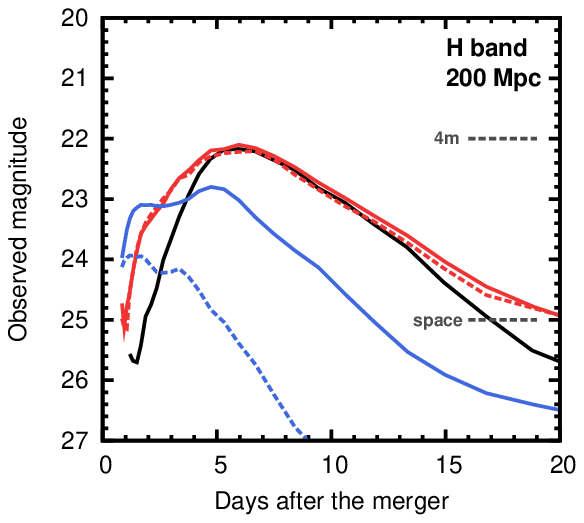} &
\includegraphics[scale=1.05]{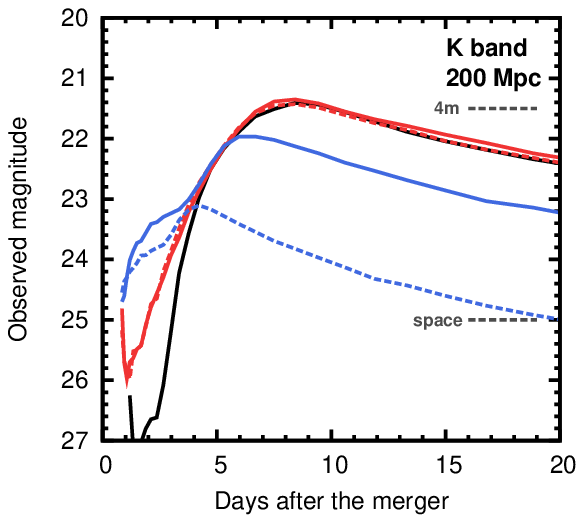}  
\end{tabular}
\caption{
Expected observed $ugrizJHK$-band light curves 
(in AB magnitude) for model NSM-all and 4 realistic models.
The distance to the NS merger event is set to be 200 Mpc.
$K$ correction is taken into account with $z = 0.05$.
Horizontal lines show typical limiting magnitudes for 
wide-field telescopes ($5 \sigma$ with 10 min exposure).
For optical wavelengths ($ugriz$ bands), 
``1 m'', ``4 m'', and ``8 m'' limits are 
taken or deduced from those of PTF \citep{law09}, 
CFHT/Megacam, and Subaru/HSC \citep{miyazaki06}, respectively.
For NIR wavelengths ($JHK$ bands), ``4 m'' and ``space'' limits 
are taken or deduced from those of Vista/VIRCAM
and the planned limits of WFIRST 
\citep{green12} and WISH \citep{yamada12}, respectively.
}
\label{fig:LCobs}
\end{center}
\end{figure*}
%%%%%%%%%%%%%%%%%%%%%%%%%%%%%%%%%%%%%%%%%%%%%%%%%%%% 

%%%%%%%%%%%%%%%%%%%%%%%%%%%%%%%%%%%%%%%%%%%%%%%%%%%% 
% Figure  %%%%%%%%%%%%%%%%%%%%%%%%%%%%%%%%%%%%%%%%%% 
%%%%%%%%%%%%%%%%%%%%%%%%%%%%%%%%%%%%%%%%%%%%%%%%%%%% 
\begin{figure}
\begin{center}
\includegraphics[scale=1.4]{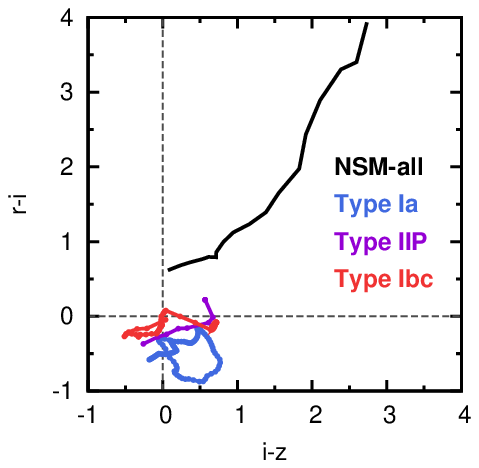} 
\caption{
Color-color diagram for the NS merger (black) compared with 
Type Ia (blue), Type IIP (purple), and Type Ic (red) SNe.
The emission from NS merger ejecta is much 
redder than that of SNe.
For SNe, we use the spectral templates by \citet{nugent02}.
All the magnitudes are in AB magnitude and 
in the rest frame (\ie no $K$ correction).
\label{fig:color}}
\end{center}
\end{figure}
%%%%%%%%%%%%%%%%%%%%%%%%%%%%%%%%%%%%%%%%%%%%%%%%%%%% 

\section{Implications for Observations}
\label{sec:implications}

\subsection{Follow-up Observations of EM Counterparts}

In this section, we discuss the detectability of 
UVOIR emission from NS merger ejecta.
Figure \ref{fig:LCobs} shows expected 
observed light curves for a NS merger event at 200 Mpc.
Model NSM-all (black) and 4 realistic models (red and blue) 
are shown.
Note that all the magnitudes in Figure \ref{fig:LCobs}
are given in AB magnitude
for the ease of comparison with different survey projects.
Horizontal lines show $5 \sigma$ limiting magnitudes 
for different sizes of telescopes with 10 min exposure time.

After the detection of GW signal, 
EM follow up observations should discover 
a new transient object from a $\sim$ 10-100 deg$^{2}$ area.
Thus, the use of wide-field telescope/camera is a natural choice
\citep[\eg][]{kelley12,nissanke13}.
For optical wavelengths, 
there are several projects using 1 m-class telescopes 
that can cover $\gsim 4$ deg$^2$ area, such as 
Palomar transient factory \citep[PTF,][]{law09,rau09}, 
La Silla-QUEST Variability Survey \citep{hadjiyska12},
and Catalina Real-Time Transient Survey \citep{drake09}.
In Figure \ref{fig:LCobs}, we show the limiting 
magnitudes deduced from \citet{law09}.
Because of the red color, the detection in 
blue wavelengths ($ug$ bands) seems difficult.
Even for the bright cases, deep observations with $>10$ min exposure
in red wavelengths ($i$ or $z$ bands) are needed.
The faint models are far below the limit of 1m-class telescopes.

For larger optical telescopes, the field of view 
tends to be smaller.
Among 4m-class telescopes, 
Canada-France-Hawaii Telescope (CFHT)/Megacam 
and the Blanco 4m telescope/DECAM 
for the Dark Energy Survey \footnote{\url{https://www.darkenergysurvey.org}}
have 3.6 deg$^2$ and 4.0 deg$^2$ field of view, respectively.
In Figure \ref{fig:LCobs}, we show the limiting magnitudes
from CFHT/Megacam
\footnote{\url{http://www.cfht.hawaii.edu/Instruments/Imaging/Megacam/generalinformation.html}}.
The bright models (red and black lines) 
are above the limits at the first 5-10 days.
Similar to 1m-class telescopes, 
observations in redder wavelengths are more efficient.
The faintest model (model H4-1314, blue dashed line) is still
below the limit of 4m-class telescopes with 10 min exposure.

To cover all the possibilities, we need 8m-class telescopes.
Among such large telescopes, only
Subaru/Hyper Suprime Cam \citep[HSC,][]{miyazaki06} and 
Large Synoptic Survey Telescope (LSST, \citealt{ivezic08,lsst09})
have a wide field of view (1.77 deg$^2$ and 9.6 deg$^2$, respectively).
We show the expected limit with Subaru/HSC.
In red optical wavelengths ($i$ or $z$ bands), 
8m-class telescope can detect even the faintest case.

In Figure \ref{fig:color}, we show a $r-i$ vs $i-z$ 
color-color diagram 
for model NSM-all compared with that of Type Ia, IIP, and Ibc SNe
\citep{nugent02}.
As clearly seen, the NS merger is significantly redder than SNe.
Thus, confusion with SNe will not be problematic.

Because of the extremely red color, follow up observations
in NIR wavelengths are also useful.
In NIR wavelengths, however, a field of view 
is usually smaller than in optical.
We plot the limit of 4.1m VISTA telescope/VIRCAM 
($\sim$0.6 deg$^2$ \citealt{dalton06})
\footnote{\url{http://www.eso.org/sci/facilities/paranal/instruments/vircam/}}.
In $J$ band, ground-based observations with 4m-class telescopes
will be able to detect a bright event in NIR wavelengths.

Observations from space seem a more promising strategy
(see also \citealt{barnes13}).
The Wide-Field Infrared Survey Telescope (WFIRST, \citealt{green12})
and Wide-field Imaging Surveyor for High-redshift (WISH, \citealt{yamada12})
are planned to perform wide-field survey in NIR wavelengths.
Their planned field of views are 0.375 deg$^2$ and 0.28 deg$^2$,
respectively.
Their typical limiting magnitude is $\sim 25$ mag with $\sim 10$ min
observations.
They will be able to detect even the faintest case (Figure \ref{fig:LCobs}).

We conclude that extensive follow-up observations 
with wide-field 4m- and 8m-class telescopes in optical
and wide-field space telescopes in NIR 
are crucial to detect the EM counterpart of GW sources.
In optical, $i$ or $z$-band observations are the most efficient.
The observations should be performed within about 5 days
(for optical) and 10 days (for NIR) from the detection of GWs.

%%%%%%%%%%%%%%%%%%%%%%%%%%%%%%%%%%%%%%%%%%%%%%%%%%%% 
% Figure  %%%%%%%%%%%%%%%%%%%%%%%%%%%%%%%%%%%%%%%%%% 
%%%%%%%%%%%%%%%%%%%%%%%%%%%%%%%%%%%%%%%%%%%%%%%%%%%% 
\begin{figure}
\begin{center}
\includegraphics[scale=1.4]{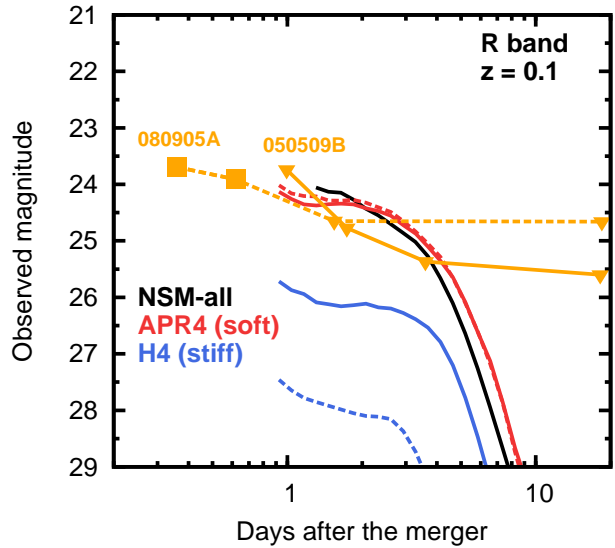} 
\caption{
Observed $R$-band light curves of the models at $z=0.1$
(in Vega magnitude, with $K$ correction)
compared with the deep detection or upper limits of 
short GRB afterglow (GRBs 050509B and 080905A).
For the afterglow data, we use $R$-band 
magnitude corrected and shifted to $z=0.1$ scale \citep{kann11}.
The squares show the detections while the triangles show
upper limits.
Deep observations of short GRB afterglow are about to 
make constraints on the emission powered by radioactive energy.
\label{fig:SGRB}}
\end{center}
\end{figure}
%%%%%%%%%%%%%%%%%%%%%%%%%%%%%%%%%%%%%%%%%%%%%%%%%%%% 

%%%%%%%%%%%%%%%%%%%%%%%%%%%%%%%%%%%%%%%%%%%%%%%%%%%% 
% Figure  %%%%%%%%%%%%%%%%%%%%%%%%%%%%%%%%%%%%%%%%%% 
%%%%%%%%%%%%%%%%%%%%%%%%%%%%%%%%%%%%%%%%%%%%%%%%%%%% 
\begin{figure*}
\begin{center}
\begin{tabular}{cc}
\includegraphics[scale=1.05]{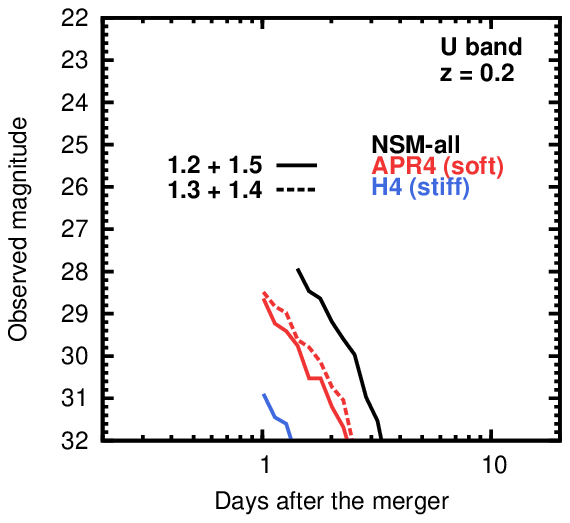} &
\includegraphics[scale=1.05]{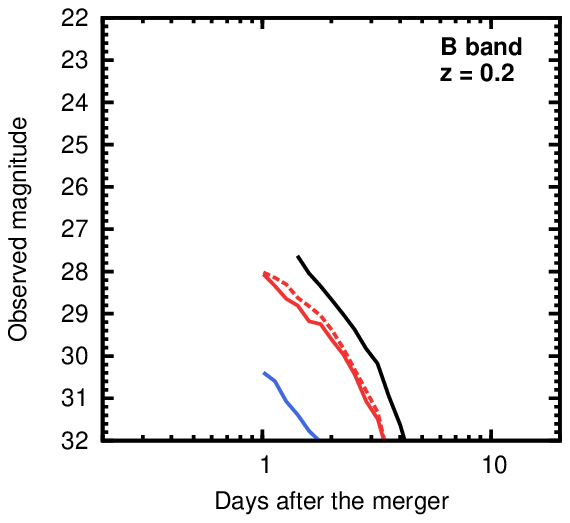} \\
\includegraphics[scale=1.05]{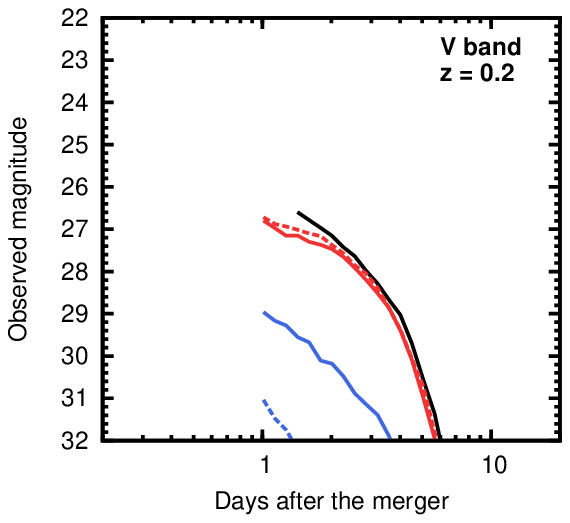} &
\includegraphics[scale=1.05]{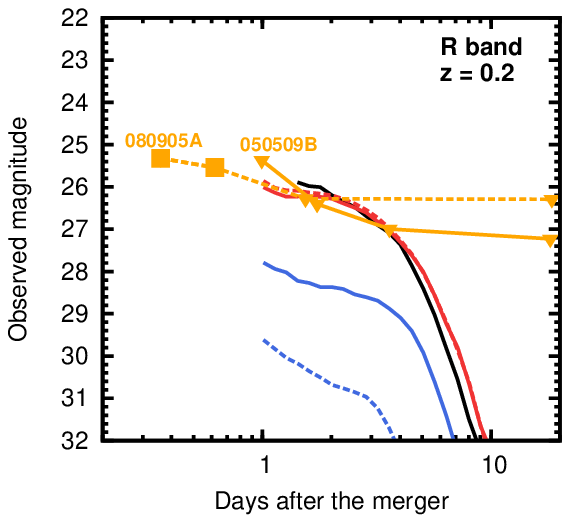} \\
\includegraphics[scale=1.05]{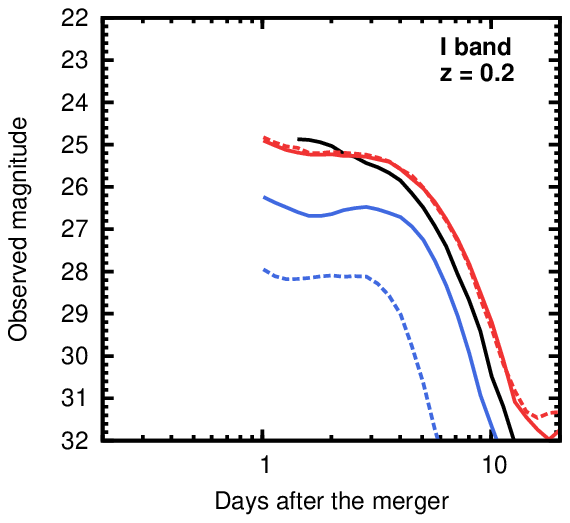} & 
\includegraphics[scale=1.05]{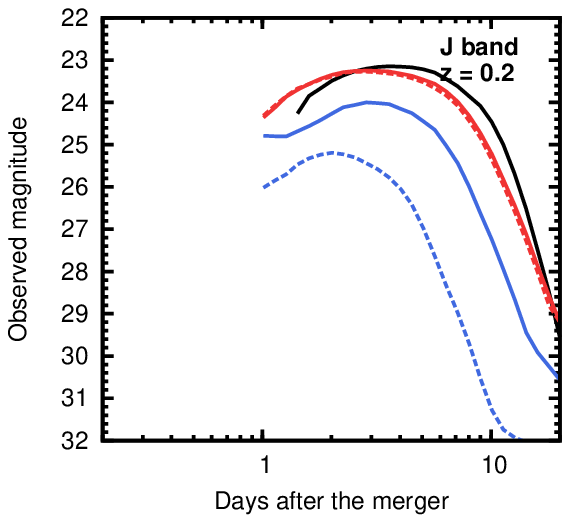} \\
\includegraphics[scale=1.05]{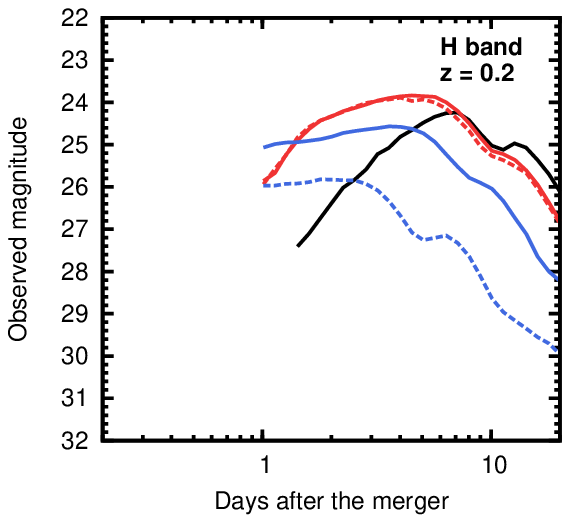} &
\includegraphics[scale=1.05]{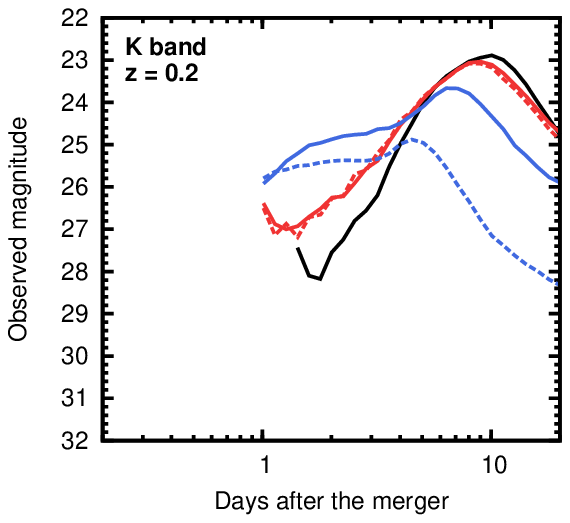}  
\end{tabular}
\caption{
Expected observed $UBVRIJHK$-band light curves 
(in Vega magnitude) of the NS merger at $z = 0.2$.
Deep follow-up observations of short GRB afterglows
will be able to detect a radioactive ``bump''. 
$K$ correction is taken into account for the models.
In the $R$-band light curve, deepest observational limits 
for short GRBs so far  
(GRBs 050509B and 080905A, \citealt{kann11}) are plotted.
The squares show the detections while the triangles show
upper limits.
Note that the original data by \citet{kann11} are 
corrected and shifted to $z=0.1$ scale, 
and we correct only the distance (to $z=0.2$) in this plot.
\label{fig:z0.2}}
\end{center}
\end{figure*}
%%%%%%%%%%%%%%%%%%%%%%%%%%%%%%%%%%%%%%%%%%%%%%%%%%%% 

\subsection{Search for Emission in Short GRB Afterglow}

Although the emission from a NS merger by radioactive decay energy
has not been detected, 
we show that it may be possible to detect the emission
as a ``bump'' in the afterglow of short GRBs
when the afterglow is faint enough.
If such a bump is detected, it proves that 
(1) nucleosynthesis involving radioactive nuclei 
takes place in the NS merger ejecta
and (2) such emission can actually be used to identify the GW sources.
Although extensive search has been performed 
for relatively nearby short GRBs,
no such an extra component has been detected 
\citep[\eg][]{fox05,hjorth05Nature,hjorth05,perley09,rowlinson10,kocevski10}.

Figure \ref{fig:SGRB} shows the model light curves at $z=0.1$.
The models (black, red and blue lines) are compared with 
the deepest observations of short GRBs so far
(GRBs 050509B and GRB 080905A) compiled by \citet{kann11}.
The data are corrected and shifted to $z=0.1$ scale 
by \citet{kann11} with an appropriate $K$-correction.
The squares show the detections while the triangles show
upper limits.
For original data, see 
\citet{gehrels05,hjorth05,castro-tirado05,bersier05,bloom06}
for GRB 050509B and \citet{rowlinson10} for GRB 080905A.
Before the correction of redshift, 
the deepest limits for GRB 050509B ($z$=0.225)
are $\sim$ 26.5 mag \citep{bersier05},
and those for 080905A (z=0.1218) are 
$\sim$ 25.0 mag \citep{rowlinson10}.

The upper limits are, in fact, very close to the 
expected light curves with $\Mej \sim 0.01 \Msun$.
We do {\it not} argue that the current deep limits 
already exclude such an ejecta mass 
because we made an assumption that the heating rate
is simply proportional to the total ejecta mass,
and there is still an uncertainty in the absolute brightness.
However, it is encouraging that 
the deepest observations actually about to make constraints on 
the emission by radioactive energy.

Figure \ref{fig:z0.2} shows the expected light curves 
at $z=0.2$ (in Vega magnitude, with appropriate $K$ corrections).
For short GRBs at $z=0.2$, $R$-band observations down to 27 (31) mag 
at $\lsim 3$ days from the burst may be able to detect
the emission powered by radioactivity 
if the bright (faint) model is the case.
As already discussed, observations in redder wavelengths 
are more efficient.
In $I$ band, observations down to 26 (29) mag may detect 
the bright (faint) case.
In NIR wavelengths, the ``bump'' can be as bright as 
23-25 mag (in Vega magnitude).
When the afterglow is faint enough not to overshine these emission,
we may be able to make constraints on the efficiency of 
mass ejection and nucleosynthesis in the NS merger ejecta
\footnote{After submission of this paper, 
possible sign of radioactive emission was reported
for short GRB 130603B at $z = 0.36$ \citep{berger13,tanvir13}.
The $H$-band magnitude is about 25-26 AB mag 
at $t \simeq 7$ days (in the rest frame).
This brightness prefers to our bright models 
(NSM-all, APR4-1215, and APR4-1314) 
rather than the faint models (H4-1215 and H4-1314).}.

\section{Conclusions}
\label{sec:conclusions}

We perform radiative transfer simulations for 
NS merger ejecta powered by radioactive energy of r-process nuclei.
This is the first simulation including all the 
r-process elements from Ga to U.
We show that the opacity in the NS merger ejecta is 
higher than previously expected by a factor of $\sim 100$
due to many bound-bound transitions of r-process elements.
A typical mass absorption coefficient is 
$\kappa \sim 10\ {\rm cm^{2}\ g^{-1}}$ 
($\kappa \sim 0.1 \ {\rm cm^{2}\ g^{-1}}$ for Fe-rich Type Ia SNe).
This is consistent with the recent results by \citet{kasen13}
and \citet{barnes13},
who computed opacity of a few lanthanoid elements 
with detailed models of these ions.
As a result of high opacity, 
the emission powered by radioactive energy
is fainter and longer than previously thought.

Spectroscopic features are important to 
identify a new source as a NS merger event.
Because of the high expansion velocity ($v \sim 0.1c$),
a single absorption line from r-process elements cannot be resolved.
In addition, by including all the r-process elements in the simulations,
spectroscopic features are almost totally smeared out.
We may recognize the NS merger event by 
their feature-less spectra with a very red color.

By using the results of numerical simulations for NS merger 
by \citet{hotokezaka13}, 
we demonstrate that NS merger with a higher mass ratio is brighter.
When a softer EOS is applied in the merger simulations,
the emission is also brighter.
This opens a new window to study the nature of the NS merger events
and EOSs by EM observations (see also \citealt{bauswein13}).

At 200 Mpc, an expected horizon for GW detection,
the expected brightness is 
$g$ = 23-27 mag, $r$ = 23 - 26 mag, $i$ = 22 - 25 mag, 
$z$ = 21 - 23 mag in optical, and 21-24 mag in NIR JHK bands (in AB magnitude).
The emission is brighter and lasts longer in redder wavelengths.
Therefore, extensive follow-up observations 
with wide-field 4m- and 8m-class telescopes in optical 
(such as CFHT/Megacam, Blanco 4m/DECAM, Subaru/HSC and LSST)
and wide-field space telescopes in NIR (\eg WFIRST and WISH)
are crucial to detect the EM counterpart of GW sources.
In optical wavelengths, 
observations in the reddest bands ($i$ or $z$ bands)
are the most efficient.
The observations should be performed within about 5-10 days
from the detection of GWs.

We show that the emission powered by radioactive energy 
can be possibly detected by deep follow-up observations of short GRB afterglow
when the afterglow is faint enough.
The current deepest limits for nearby short GRBs are already
very close to the expected brightness.
When our bright (faint) model is the case, 
observations down to $R = 27$ (31) mag, 
$I = 26$ (29) mag, or 24 (26) mag in NIR JHK bands (in Vega magnitude)
for short GRBs at $z=0.2$ will be able to detect the ``bump''.
If such emission is detected, it provides evidence that 
(1) nucleosynthesis involving radioactive nuclei 
takes place in the NS merger ejecta
and (2) such emission can actually be used to identify the GW sources.

%\clearpage

\acknowledgments
The authors thank Yuichiro Sekiguchi, Masaru Shibata, Kenta Kiuchi, 
Keiichi Maeda, and Koutarou Kyutoku for fruitful discussion 
that launched this work.
We also thank Sergei Blinnikov, Dan Kasen, Markus Kromer, Leon Lucy,  
Stuart Sim, and Elena Sorokina for providing their results of 
radiative transfer simulations, and Ken Nomoto for providing W7 model.
MT thanks Akimasa Kataoka, Shinya Wanajo, and Kunihito Ioka 
for valuable discussion.
We have made extensive use of NIST database for atomic data, 
and VALD database \citep{piskunov95,ryabchikova97,kupka99,kupka00}
for line lists.
Atomic data compiled in the DREAM data base \citep{biemont99} 
were extracted via VALD.
A large part of numerical simulations presented in this paper 
were carried out with Cray XC30 at Center for Computational Astrophysics, 
National Astronomical Observatory of Japan.
This research has been supported 
by the Grant-in-Aid for Scientific Research of the 
Japan Society for the Promotion of Science (JSPS, 24740117)
and Grant-in-Aid for Scientific Research on Innovative Areas
of the Ministry of Education, Culture, Sports, Science and Technology 
(MEXT, 25103515).

%\bibliography{../../Reference/reference}

%%%%%%%%%%%%%%%%%%%%%%%%%%%%%%%%%%%%%%%%%%%%%%%%%%%% 
% Appendix  %%%%%%%%%%%%%%%%%%%%%%%%%%%%%%%%%%%%%%%% 
%%%%%%%%%%%%%%%%%%%%%%%%%%%%%%%%%%%%%%%%%%%%%%%%%%%% 

\appendix

\section{Appendix A. Test calculations}

%%%%%%%%%%%%%%%%%%%%%%%%%%%%%%%%%%%%%%%%%%%%%%%%%%%% 
% Figure  %%%%%%%%%%%%%%%%%%%%%%%%%%%%%%%%%%%%%%%%%% 
%%%%%%%%%%%%%%%%%%%%%%%%%%%%%%%%%%%%%%%%%%%%%%%%%%%% 
\begin{figure*}
\begin{center}
\begin{tabular}{cc}
\includegraphics[scale=1.3]{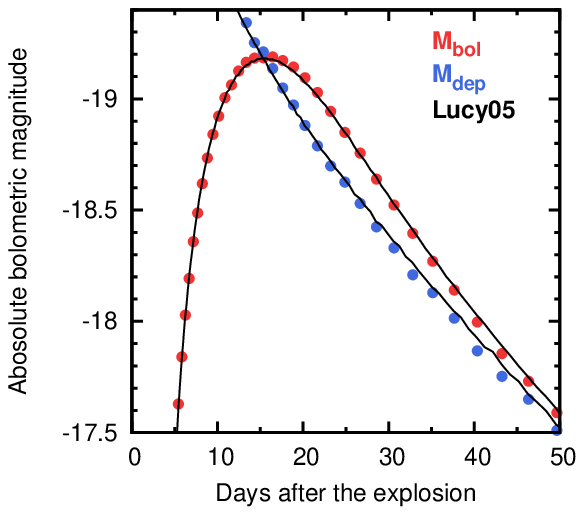} &
\includegraphics[scale=1.3]{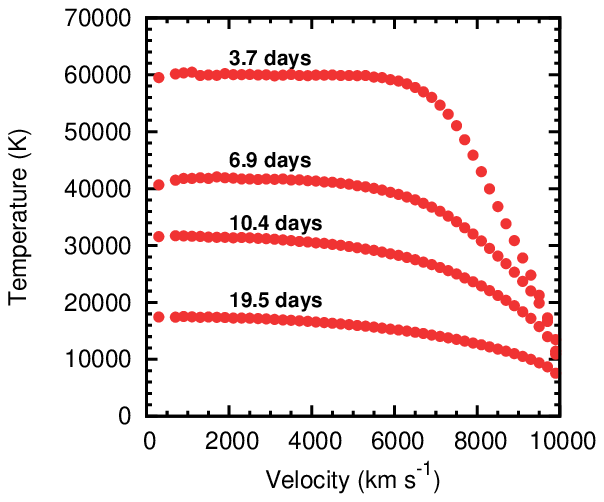}
\end{tabular}
\caption{
Test calculations with the simple Type Ia SN model by \citet{lucy05}
under gray approximation for UVOIR transfer.
({\it Left}) The bolometric luminosity (red)
and deposited luminosity (blue) in absolute magnitude 
calculated with the new code.
The solid lines show the corresponding results by \citet{lucy05}.
({\it Right}) Temperature structure as a function of velocity
for different epochs.
\label{fig:gray}}
\end{center}
\end{figure*}
%%%%%%%%%%%%%%%%%%%%%%%%%%%%%%%%%%%%%%%%%%%%%%%%%%%% 

\subsection{A.1. Gray transfer}

First we apply our new code for a simple Type Ia SN model
introduced by \citet{lucy05}.
The model has a uniform density distribution 
up to the maximum ejecta velocity of 10,000 \kms.
The total ejecta mass is 1.39 $\Msun$.
The model includes 0.65 $\Msun$ of \Nifs.
The distribution of \Nifs\ is assumed to be constant 
at $M_r < 0.5 \Msun$, while it drops linearly to zero
at $M_r = 0.75 \Msun$.

UVOIR transfer is performed under the gray approximation
with $\kappa = 0.1 \ {\rm cm^{2}\ g^{-1}}$.
The major difference between our code and that by \citet{lucy05}
is the $\gamma$-ray transfer.
\citet{lucy05} solves multi-energy $\gamma$-ray transport by
taking into account Compton scattering and photoelectric absorption
while our code adopt the gray approximation.
We use an effective mass absorption coefficient of 
$\kappa = 0.027 \ {\rm cm^{2}\ g^{-1}}$,
which is known to reproduce the results of multi-energy transport
and the observed light curves of Type Ia SNe
\citep{colgate80,sutherland84,maeda06gamma}.

Our calculations (dots) and those by \citet[lines]{lucy05}
show fairly good agreement (left panel of Figure \ref{fig:gray}).
This indicates that our MC radiation solver works properly.
Although we do not require a temperature structure for this calculation,
\citet{kasen06} showed the temperature structure  
calculated with their 3D MC code.
The temperature computed by our code (right panel of Figure \ref{fig:gray})  
shows an excellent match with that by \citet{kasen06}.
It indicates that
the temperature estimate in our code also works reasonably
(Section \ref{sec:temperature}).

\subsection{A.2. Multi-frequency transfer}

%%%%%%%%%%%%%%%%%%%%%%%%%%%%%%%%%%%%%%%%%%%%%%%%%%%% 
% Figure  %%%%%%%%%%%%%%%%%%%%%%%%%%%%%%%%%%%%%%%%%% 
%%%%%%%%%%%%%%%%%%%%%%%%%%%%%%%%%%%%%%%%%%%%%%%%%%%% 
\begin{figure*}
\begin{center}
\begin{tabular}{ccc}
\includegraphics[scale=0.8]{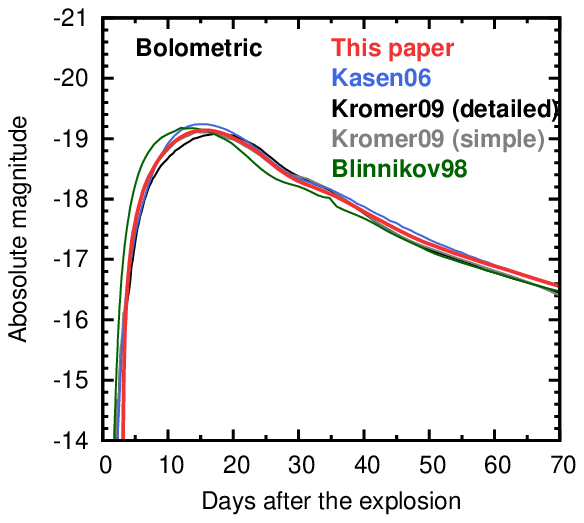} &
\includegraphics[scale=0.8]{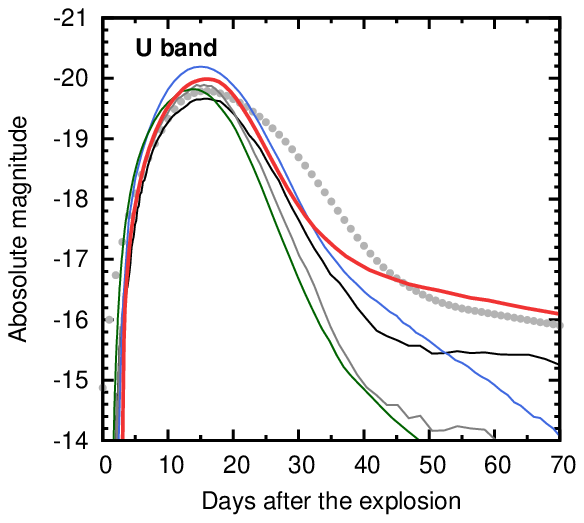} &
\includegraphics[scale=0.8]{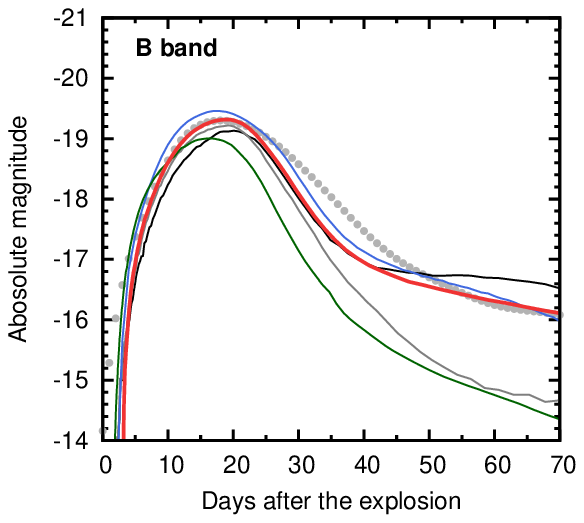} \\
\includegraphics[scale=0.8]{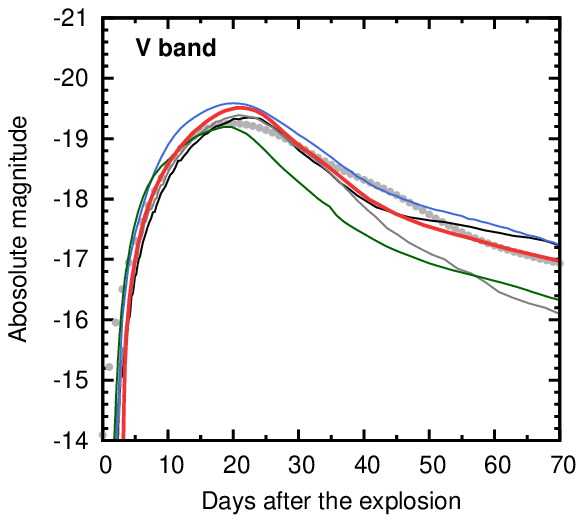} &
\includegraphics[scale=0.8]{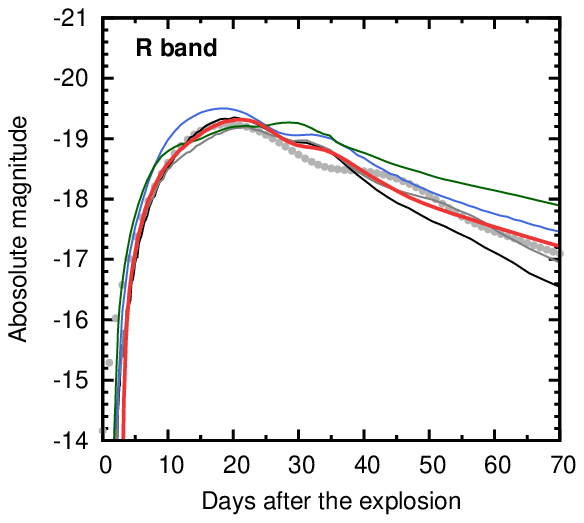} &
\includegraphics[scale=0.8]{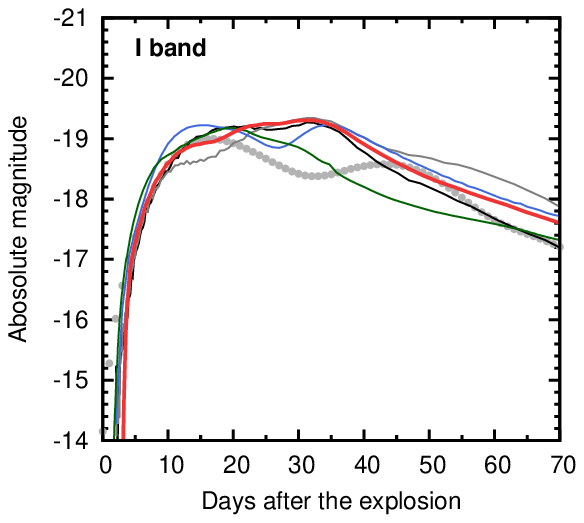} \\
\includegraphics[scale=0.8]{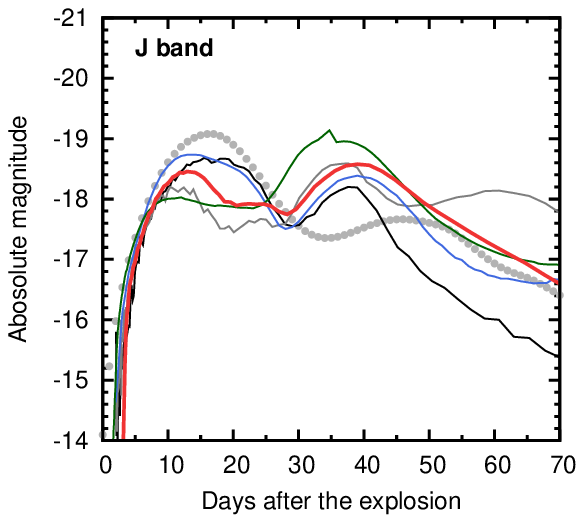} &
\includegraphics[scale=0.8]{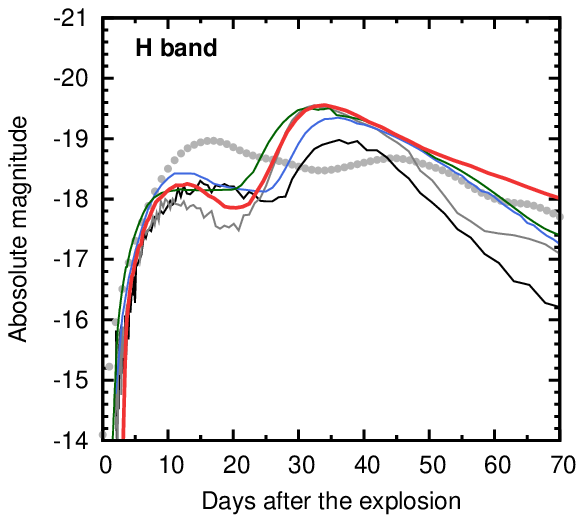} &
\includegraphics[scale=0.8]{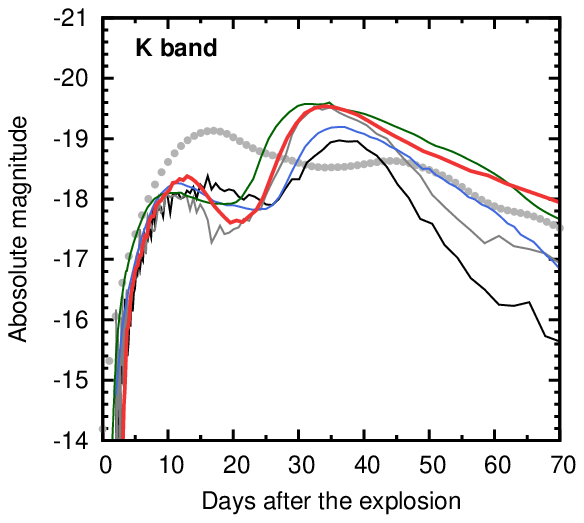} \\
\end{tabular}
\caption{
Comparison of light curves for W7 model with different numerical codes
(lines):
3D MC codes by \citet[blue]{kasen06}, \citet[black and gray]{kromer09}, 
and ours (red),
and 1D radiation hydrodynamic code by \citet[green]{blinnikov98}.
They are also compared with the template light curves of Type Ia SN
by \citet[gray dots]{hsiao07}.
For the template light curve, the rise time in $B$ band 
is assumed to be 18 days.
\label{fig:LC_W7}}
\end{center}
\end{figure*}
%%%%%%%%%%%%%%%%%%%%%%%%%%%%%%%%%%%%%%%%%%%%%%%%%%%% 

%%%%%%%%%%%%%%%%%%%%%%%%%%%%%%%%%%%%%%%%%%%%%%%%%%%% 
% Figure  %%%%%%%%%%%%%%%%%%%%%%%%%%%%%%%%%%%%%%%%%% 
%%%%%%%%%%%%%%%%%%%%%%%%%%%%%%%%%%%%%%%%%%%%%%%%%%%% 
\begin{figure}
\begin{center}
\includegraphics[scale=1.4]{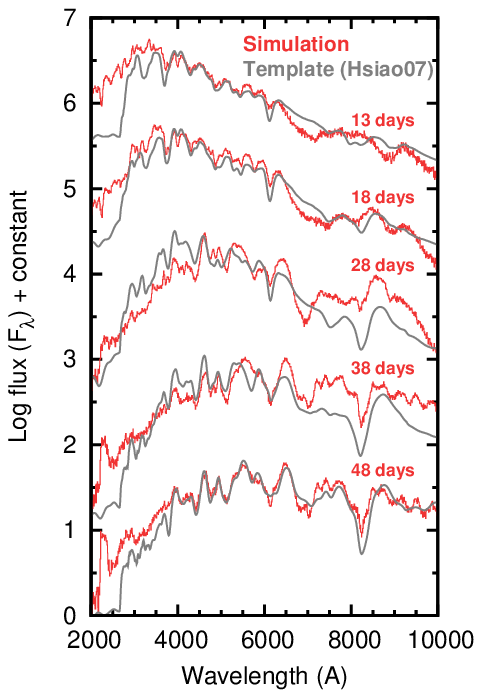} 
\caption{
Computed time series of optical spectra for W7 model
(red) compared with the spectral template of Type Ia SN 
by \citet[gray]{hsiao07}.
The epoch shown in the panel is the time after the explosion.
Observational template at the corresponding epoch are overplotted
assuming the rise time in $B$ band to be 18 days.
The flux is arbitrarily scaled.
\label{fig:spec_W7}}
\end{center}
\end{figure}
%%%%%%%%%%%%%%%%%%%%%%%%%%%%%%%%%%%%%%%%%%%%%%%%%%%% 

Next we apply our code for W7 model of Type Ia SN
\citep{nomoto84} with multi-frequency UVOIR transfer.
This model has been used to test 
multi-frequency radiative transfer codes for SN ejecta.
The model has a stratified abundance distribution;
stable Fe-group layer ($M_r \lsim 0.2 \Msun$), 
\Nifs\ layer ($M_r \simeq 0.2-0.8 \Msun$, 
the total mass of \Nifs\ is about 0.6 $\Msun$), 
Si and S-rich intermediate-mass element layer ($M_r \simeq 0.8-1.1 \Msun$),
and O-rich layer ($M_r \gsim 1.1 \Msun$) from the center to the surface.
Such a stratified distribution is also supported from
observations \citep[see \eg][]{stehle05,mazzali07Ia,mazzali0804eo,tanaka11}.

Figure \ref{fig:LC_W7} shows comparisons of the bolometric 
and monochromatic light curves for W7 with different numerical codes.
Models are also compared with the template light curves of Type Ia SN
by \citet[gray dots]{hsiao07}.
The red lines show our calculations.
The blue lines show the results with the 3D MC code ({\sc SEDONA}) 
by \citet{kasen06}.
This code adopts a big line list (including about $4 \times 10^7$ lines) 
by \citet{kurucz93}.
The black and gray lines show the results with the 3D MC code ({\sc ARTIS})
by \citet{sim07} and \citet{kromer09}.
The black line (``detailed'') shows the calculations with 
detailed ionization treatment
by taking into account non-LTE effects and with a big bound-bound line list
(including about $8 \times 10^6$ lines by \citealt{kurucz06}).
The gray line ("simple") shows the calculations under LTE assumption with 
a moderate line list by \citet{kurucz95}, which our code also adopts.
Thus, their simple case is more similar to our code.
The green lines show the results with the 1D radiation hydrodynamic
code by \citet{blinnikov98} \citet{blinnikov00}.

Given the complexity of the problem, 
the overall agreement among different codes is reasonable.
The agreement in the bolometric luminosity is almost perfect.
Generally, the agreement is better in optical wavelengths
than in NIR wavelengths.
The $U$-band light curves starts to differ at $> 30$ days after the explosion.
Our code provides the best match with the observations at later epochs.
The computed $B$, $V$, and $R$-band light curves agree with each other 
quite well, and they are also consistent with observations.
The computed $I$-band light curves also agree among different codes,
but they cannot reproduce a clear two peaks seen in the observations.
This may be related to the treatment (absorptive or scattering) 
of the \ion{Ca}{ii} IR triplet line \citep{kasen06}.
All the computed $J$, $H$, and $K$-band light curves show 
the two peaks, which are roughly consistent with the observations.
With a closer look, computed light curves tend to show a fainter first
peak and a brighter second peak than the observations.
\citet{kromer09} demonstrated the number of bound-bound transitions 
included in the simulation mostly affects the NIR light curves
(black and gray lines in Figure \ref{fig:LC_W7}).
It is encouraging that with a bigger line list, 
the first (second) peak becomes brighter (fainter),
which is closer to the observations.
See also \citet{sim10} for a similar comparison and discussion.

Figure \ref{fig:spec_W7} shows the computed optical spectra 
at different epochs compared with the template spectra \citep{hsiao07}
in the corresponding phase.
This also shows a reasonable agreement.
As expected from the comparison of the light curves
(Figure \ref{fig:LC_W7}), 
the agreement of the overall color is not perfect at some epochs.
Nevertheless, the strong absorption lines, such as those of 
\ion{O}{i}, \ion{Si}{ii}, \ion{S}{ii}, \ion{Ca}{ii}, and \ion{Fe}{ii},
are produced at the right positions in the simulated spectra.
This indicates that our code reasonably computes ionization states
and opacity.

From these comparisons, we conclude that 
our new code works for multi-frequency UVOIR transfer
with the Type Ia SN model.
Simulations for NS merger ejecta are different from those for Type Ia SNe
mostly in that the NS merger ejecta consist of heavier elements.
However, we can consistently use the same Saha equation solver 
and the same format of the line list.
Therefore, the simulations for NS merger do not require 
any computational technique that is not used
in the simulation for Type Ia SNe.

%%%%%%%%%%%%%%%%%%%%%%%%%%%%%%%%%%%%%%%%%%%%%%%%%%%% 
% Figure  %%%%%%%%%%%%%%%%%%%%%%%%%%%%%%%%%%%%%%%%%% 
%%%%%%%%%%%%%%%%%%%%%%%%%%%%%%%%%%%%%%%%%%%%%%%%%%%% 
\begin{figure}
\begin{center}
\includegraphics[scale=1.4]{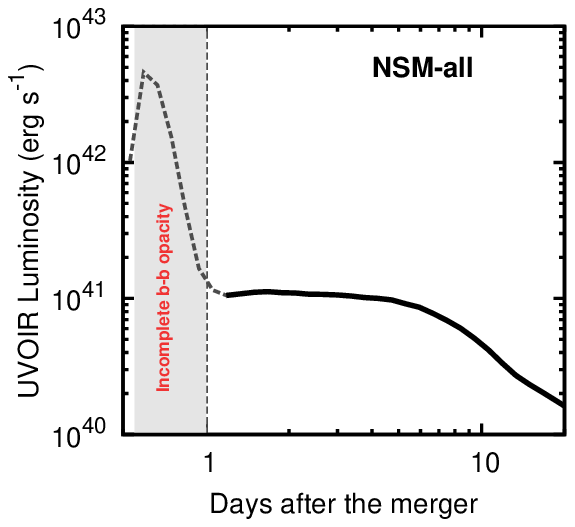} 
\includegraphics[scale=1.4]{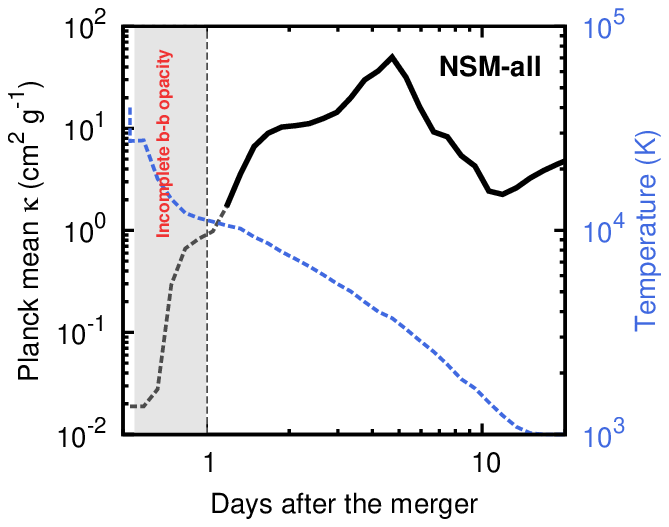} 
\caption{
({\it Upper}) Bolometric light curve of model NSM-all.
({\it Lower}) Planck-mean opacity (black) 
and temperature (blue) at $v=0.1 c$. 
Gray hatched area shows the epoch when 
the temperature at $v = 0.1 c$ is $T \gsim 10000$ K.
At such early epochs, our line list is not applicable, 
and the opacity is extremely low.
This low opacity makes 
an unphysical early peak in the light curve.
\label{fig:NSM_early}}
\end{center}
\end{figure}
%%%%%%%%%%%%%%%%%%%%%%%%%%%%%%%%%%%%%%%%%%%%%%%%%%%% 

%%%%%%%%%%%%%%%%%%%%%%%%%%%%%%%%%%%%%%%%%%%%%%%%%%%% 
% Figure  %%%%%%%%%%%%%%%%%%%%%%%%%%%%%%%%%%%%%%%%%% 
%%%%%%%%%%%%%%%%%%%%%%%%%%%%%%%%%%%%%%%%%%%%%%%%%%%% 
\begin{figure}
\begin{center}
\includegraphics[scale=1.3]{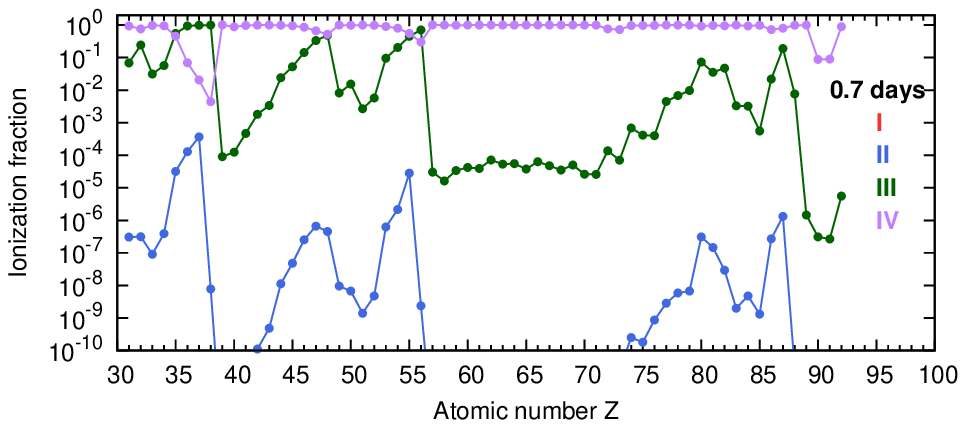} 
\includegraphics[scale=1.3]{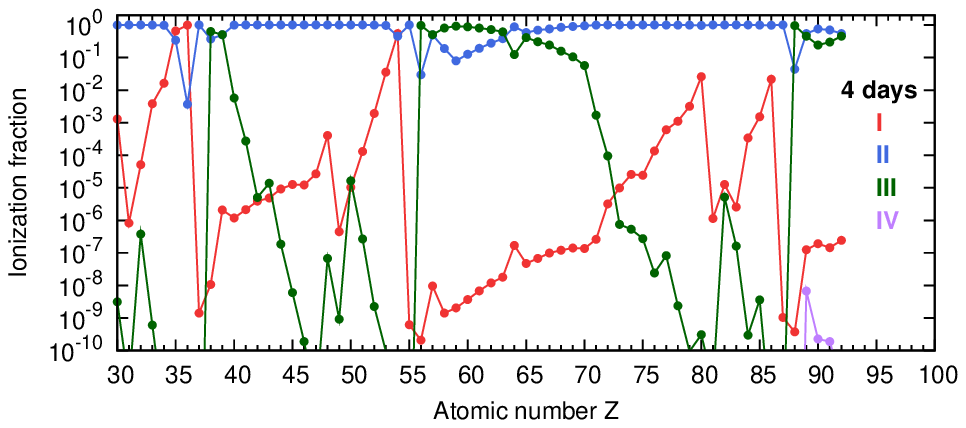} 
\caption{
Ionization states for different elements at $v=0.1c$
(in model NS-all) at $t = 0.7$ days (upper) and 4 days (lower) 
after the merger.
It is shown that at $t=0.7$ days, most of elements 
are triply ionized, which our line list does not cover
(see Figure \ref{fig:nline}).
At $t=4$ days, dominant ionization states are singly or 
doubly ionized ions. 
\label{fig:ionization}}
\end{center}
\end{figure}
%%%%%%%%%%%%%%%%%%%%%%%%%%%%%%%%%%%%%%%%%%%%%%%%%%%% 

\section{Appendix B. Limitation of Our Line List}

In the figures in the main text, we show our results of 
multi-frequency transfer for NS mergers only at $t \gsim 1$ day.
Here we show the whole range of our results and discuss 
the limitation of our line list.
The upper panel of Figure \ref{fig:NSM_early} shows the 
whole range of the computed bolometric light curve for
model NSM-all.
As clearly seen, the luminosity at $t < 1$ day is extremely
high, reaching $\sim 5 \times 10^{43}\ {\rm erg\ s^{-1}}$.
The lower panel of Figure \ref{fig:NSM_early} shows 
the Planck-mean mass absorption coefficient at $v = 0.1c$
in model NSM-all.
The opacity at $t < 1$ day is extremely low, starting from 
$\sim 2 \times 10^{-2}\ {\rm cm^2\ g^{-1}}$,
which is found to be dominated by the electron scattering.

The reason of this low opacity is the incompleteness of 
our line list.
As shown in Figure \ref{fig:nline}, our line list does not 
include the bound-bound data of triply ionized ions for the 
heavy elements with $Z \ge 31$.
However, the ejecta are dominated by triply ionized ions 
at $t < 1$ day.
Figure \ref{fig:ionization} shows the ionization fractions 
for different elements in the ejecta.
At $t$ = 0.7 day, triply ionized ions (purple) are dominated
over the lower ionization states.
As a result, the bound-bound opacity at such early epochs 
cannot be evaluated correctly.

This situation changes at later epochs.
When the temperature is lower than 10,000 K (blue line in 
Figure \ref{fig:NSM_early}), 
the dominant ionization states are no more triply ionized ions.
The lower panel of Figure \ref{fig:ionization} shows the 
ionization at $t=4$ day.
At this epoch, the ejecta are dominated by singly and doubly ionized ions.
Therefore, we use only the results at the epochs when 
the temperature at $v=v_{\rm ch}$ is below 10,000 K.

One concern is a possible effect of the very low opacity 
to the later epoch.
However, as shown in Figure \ref{fig:NSM-all}, 
the results of the multi-frequency transfer at $t > 1$ day 
closely follow the light curve with the gray transfer with 
$\kappa = 10\ {\rm cm^2\ g^{-1}}$.
This is consistent with the expectation 
from the Planck-mean opacity at $t>1 $ day 
in multi-frequency transfer (Figure \ref{fig:NSM_early}).
Thus, we conclude that the light curve at $t > 1$ day is not 
significantly affected by the low opacity at earlier epochs.

It is noted that, for the case of SNe, 
we do not encounter a similar problem
primarily because
the bound-bound data include triply ionized ions at $Z \le 30$.
In addition to this, 
there is an important difference in the ionization states
in the SN ejecta and NS merger ejecta.
Since the mean atomic mass is heavier in the NS merger ejecta,
the number density of ions is smaller in the NS merger ejecta 
than in Type Ia SN ejecta.
As a result, the number density of free electrons is also smaller
in the NS merger ejecta.
Thus, for a given temperature and a density, 
the ionization states in the NS merger ejecta 
are higher than those in Type Ia SN ejecta.
Therefore, simulations for NS mergers
more easily encounter the incompleteness 
in the bound-bound data of highly-ionized ions.

\end{document}